%% file: Main_CCS.tex
\documentclass[sigconf,authorversion,nonacm]{acmart}

\usepackage{hologo}
\usepackage{xspace}
\usepackage{makecell}
\usepackage[ruled,linesnumbered]{algorithm2e}
\usepackage{hologo}
\usepackage{hyperref}
\usepackage{bbding}
\usepackage{pifont}
\usepackage{wasysym}
\usepackage{utfsym}
\usepackage{pifont}
\usepackage{array}
\usepackage{url}
\usepackage[tight,footnotesize]{subfigure}

\usepackage{cleveref}   
\usepackage{booktabs}
\usepackage{graphicx}
\usepackage{booktabs}
\usepackage{multirow}
\usepackage{longtable}
\usepackage{rotating} 
\usepackage{booktabs} 
\usepackage{lscape} 

\AtBeginDocument{%
  \providecommand\BibTeX{{%
    \normalfont B\kern-0.5em{\scshape i\kern-0.25em b}\kern-0.8em\TeX}}}

\begin{document}



\title{PARIS: A Practical, Adaptive Trace-Fetching and Real-Time Malicious Behavior Detection System}
\keywords{Malware Analysis, Behavior Recognition, Advanced Persistent Threat (APT), Real-time Detection, Feature Engineering}
\author{Jian Wang}
\authornote{Equal Contribution}
\affiliation{%
  \institution{College of Computer Science and Technology, Zhejiang University}
  \city{Hangzhou}
  \postcode{310027}
  \country{China}
}
\email{wangjian1998@zju.edu.cn}

\author{Lingzhi Wang}
\authornotemark[1] 
\affiliation{%
  \institution{Department of Computer Science, Northwestern University}
  \city{Evanston}
  \state{Illinois}
  \postcode{60208}
  \country{USA}
}
\email{lingzhiwang2025@u.northwestern.edu}

\author{Husheng Yu}
\affiliation{%
  \institution{College of Computer Science and Technology, Zhejiang University}
  \city{Hangzhou}
  \postcode{310027}
  \country{China}
}
\email{yhsheng@zju.edu.cn}

\author{Xiangmin Shen}
\affiliation{%
  \institution{Department of Computer Science, Northwestern University}
  \city{Evanston}
  \state{Illinois}
  \postcode{60208}
  \country{USA}
}
\email{XiangminShen2019@u.northwestern.edu}

\author{Yan Chen}
\affiliation{%
  \institution{Department of Computer Science, Northwestern University}
  \city{Evanston}
  \state{Illinois}
  \postcode{60208}
  \country{USA}
}
\email{ychen@northwestern.edu}


\begin{abstract}
  \input{0_Abstract}
\end{abstract}



\keywords{Malware Analysis, Behavior Recognition, Advanced Persistent Threat (APT), Real-time Detection, Feature Engineering}


\settopmatter{printfolios=true}

\maketitle

\input{1_intro}
\input{2_background}

\input{3_method}
\input{4_implementation}
\input{5_experiment}
\input{6_discussion}
\input{7_related}
\input{8_conclusion}

\bibliographystyle{ACM-Reference-Format}
\bibliography{Reference}

\appendix
\input{appendix}

\end{document}

%% file: 0_abstract.tex
The escalating sophistication of cyber-attacks and the widespread utilization of stealth tactics have led to significant security threats globally. 
Nevertheless, the existing static detection methods exhibit limited coverage, and traditional dynamic monitoring approaches encounter challenges in bypassing evasion techniques.
Thus, it has become imperative to implement nuanced and dynamic analysis to achieve precise behavior detection in real time.
There are two pressing concerns associated with current dynamic malware behavior detection solutions. Firstly, the collection and processing of data entail a significant amount of overhead, making it challenging to be employed for real-time detection on the end host. Secondly, these approaches tend to treat malware as a singular entity, thereby overlooking varied behaviors within one instance.

To fill these gaps, we propose PARIS, an adaptive trace fetching, lightweight, real-time malicious behavior detection system. 
Specifically, we monitor malicious behavior with Event Tracing for Windows (ETW) and learn to selectively collect maliciousness-related APIs or call stacks, significantly reducing the data collection overhead.
As a result, we can monitor a wider range of APIs and detect more intricate attack behavior.

We implemented a prototype of PARIS and evaluated the system overhead, the accuracy of comparative behavior recognition, and the impact of different models and parameters.
The result demonstrates that PARIS can reduce over 98.8\% of data compared to the raw ETW trace and hence decreases the overhead on the host in terms of memory, bandwidth, and CPU usage with a similar detection accuracy to the baselines that suffer from the high overhead.
Furthermore, a breakdown evaluation shows that 80\% of the memory and bandwidth savings and a complete reduction in CPU usage can be attributed to our adaptive trace-fetching collector. 

%% file: 1_intro.tex
\section{Introduction}\label{introduction}
The exponential growth and ubiquitous use of the internet bring a significant increase in complex cyber attacks, which pose significant security risks on a global scale and have resulted in substantial financial losses~\cite{fireeye, sony, APT1}.
Various types of malware play an essential role in these attacks, with attackers often using their built-in malicious behavior to conduct cyber attacks, remotely monitoring and controlling the victim's host~\cite{darkcomet,Xtremerat}.
For example, the DarkComet appeared in the conflict of Syria and is used by criminals to circumvent government censorship and conduct Internet surveillance~\cite{darkcomet}. Moreover, the Xtreme was used in APT attacks against Middle Eastern countries~\cite{Xtremerat}. 

After reviewing more than 500 white papers~\cite{whitepaper} of over 50 malware families~\cite{rat2,rat3}, we found that most APT attacks target Windows systems and exhibit similar malicious behaviors~\cite{whitepaper,threatpost}.
Most of these behaviors belong to the post-compromise stage, including keylogging, remote desktop, remote shell, file system management, recording, etc.
Thus, detecting and analyzing malware behavior on Windows is a significant task.
Lastly, some routine activities are necessary for both benign software and malware.
Besides, because a large amount of malware is constructed by inserting malicious components into benign software~\cite{fan2018android}, the legitimate part might evade the detection systems if the attackers hide their malicious behaviors temporally.
Thus, detecting similar malware behaviors will be more efficient than merely detecting malware.

Abundant work~\cite{li2022novel,venkatraman2019hybrid,hemalatha2021efficient,alzaylaee2020dl,ye2007imds,preda2008semantics,maniriho2023api} have been proposed for malware detection. 
Static program analysis-based malware detection~\cite{alazab2011zero}, as previous researched, could be easily bypassed  by obfuscation~\cite{sung2004static, kendall2007practical, sung2004static} and polymorphism~\cite{bazrafshan2013survey}.
Besides, utilizing local malware analysis models take the risk that they may be hacked by attackers~\cite{corona2013adversarial} while uploading malware samples to server-side models occupy lots of bandwidth resources~\cite {yang2020ratscope}. Thus, static analysis is not suitable for real-time detection on the end host.

Dynamic analysis-based detection partially solves the obfuscation~\cite{cho2014malware} and uploading problem~\cite{yang2020ratscope} by dynamically collecting malware's run-time features and analyzing their dynamic behavior~\cite{amer2020contextual,sami2010malware,zhang2014semantics,}. 
Typical run-time features including API call sequences~\cite{ki2015novel,cono2020designing,abed2015applying,bose2008behavioral,ji2016multi,wang2019mobile}, OP code~\cite{tobiyama2016malware}, system calls~\cite{milajerdi2019holmes,han2020unicorn,ahmed2020system} and audit logs~\cite{zeng2021watson}.
However, collecting these features introduces a non-negligible overhead. 
Many previous studies rely on sandbox~ \cite{tobiyama2016malware,sandbox1,sandbox2} or virtual machine~\cite{ki2015novel} for data collection. This type of work generally has high overheads and does not allow for real-time collection and detection\cite{zhang2023building,wan2019practical} on the client side.
In addition, they may be detected by malware to evade such monitoring.
Even collection techniques that don't require virtualization, such as API hooks, will typically consume 15\% of the system's resources as overhead~\cite{ding2022seqtrace,ki2015novel,tian2010differentiating}. 
In addition, for performance reasons, this type of work usually analyses only a limited number of APIs \cite{lopez2017survey}. For example, Hsiao et al. focus on only 22 APIs \cite{hsiao2020hardware}, while Sung et al. focus only on the APIs in a specific dynamic link library (kernel32.dll) \cite{sun2006api}.

A pragmatic concern in designing practical detection systems exists regarding the balance between overhead and precision. 
Collecting and analyzing more data brings more overhead.
In contrast, analyzing less data may come at the cost of accuracy. 
Especially when it comes to fine-grained semantic recognition, which is particularly important in understanding attackers' intentions in cyber attacks.
Implementing a real-time, low-overhead, yet accurate malware detection system remains an open research problem. 

Event Tracing for Windows (ETW)~\cite{ETW} as a Microsoft native auditing logging tool is widely used in Windows for log collection~\cite{ahmed2021peeler} with advantages such as stability, instrumentation-free, and relatively low overhead. 
Nevertheless, ETW has many modules and optional data for collection, and enabling too many options will introduce unacceptable overheads.
To ensure low overhead, several previous works based on ETW have only gathered high-level event information (e.g., process and file events) \cite{rana2022automated,ahmed2021peeler}, disregarding a significant amount of low-level call stack data leads to inadequate semantic identification performance.
But even then, many of these events still need to be cropped for real-time forensics.\cite{zhu2021general}.
At the same time, some efforts choose fine-grained call stack information as a data source.
For instance, RATScope~\cite{yang2020ratscope} utilizes the complete set of call stack data from ETW to detect malware behavior, resulting in an excessive workload that limits its online execution capability.
Conversely, CONAN~\cite{xiong2020conan} relies solely on top-level APIs, compromising its detection accuracy.
Thus, to attain precision and effectiveness in detection, it is imperative to select pertinent data meticulously. 

In Summary, there are several knotty challenges in designing practical, adaptive trace-fetching and real-time malicious behavior detection systems:

\textbf{C1: Collecting (along with parsing and detecting) fine-grained API calls usually brings a huge overhead and delay.}
In the previous detection work based on API calls, whether it is hook ~\cite{dahse2014simulation,qu2016appshield}, sandbox~\cite{xing2020research,li2022novel}, or auditing tools~\cite{bates2015trustworthy,gehani2012spade,king2005enriching}, it would bring a significant overhead, making it impossible to run real-time at low cost.
Even ETW-based approaches~\cite{xiong2020conan,ma2015accurate,ahmed2021peeler,wei2021deephunter} can only handle some coarse-grained security-related events such as processes, files, and sockets for efficiency consideration, which makes them only able to diagnose attacks but have no knowledge of behaviors.
To the best of our knowledge, how to efficiently handle fine-grained ETW data (e.g., system call stacks) is still an unsolved problem.

\textbf{C2: Analyzing malware behaviors accurately is challenging.}
As mentioned, identifying malicious behavior is more significant for detecting advanced cyber attacks such as APT.
However, it is difficult to evaluate the behavior detection capability of dynamic malware detection methods due to the difficulty in determining the exact number and time of the behaviors in the trace data.

\textbf{C3: Domain expertise is usually required when analyzing API calls.}
The astronomical number of API calls makes it hard to run any machine-learning algorithm.
To control the complexity of the machine learning model, previous work may use the prior knowledge about the APIs to classify them into a few categories ~\cite{ki2015novel,amer2020contextual, ahmed2009using}, or only use a specific subset of APIs based on prior knowledge~\cite{kim2018ntmaldetect,fan2018android,ahmed2009using}, which introduce significant limitations and biases to the detection.
Realizing the automated analysis and selection of API functions without introducing additional knowledge is still a great challenge.

\begin{figure}[h!]
    \centering
    \includegraphics[width=0.46\textwidth]{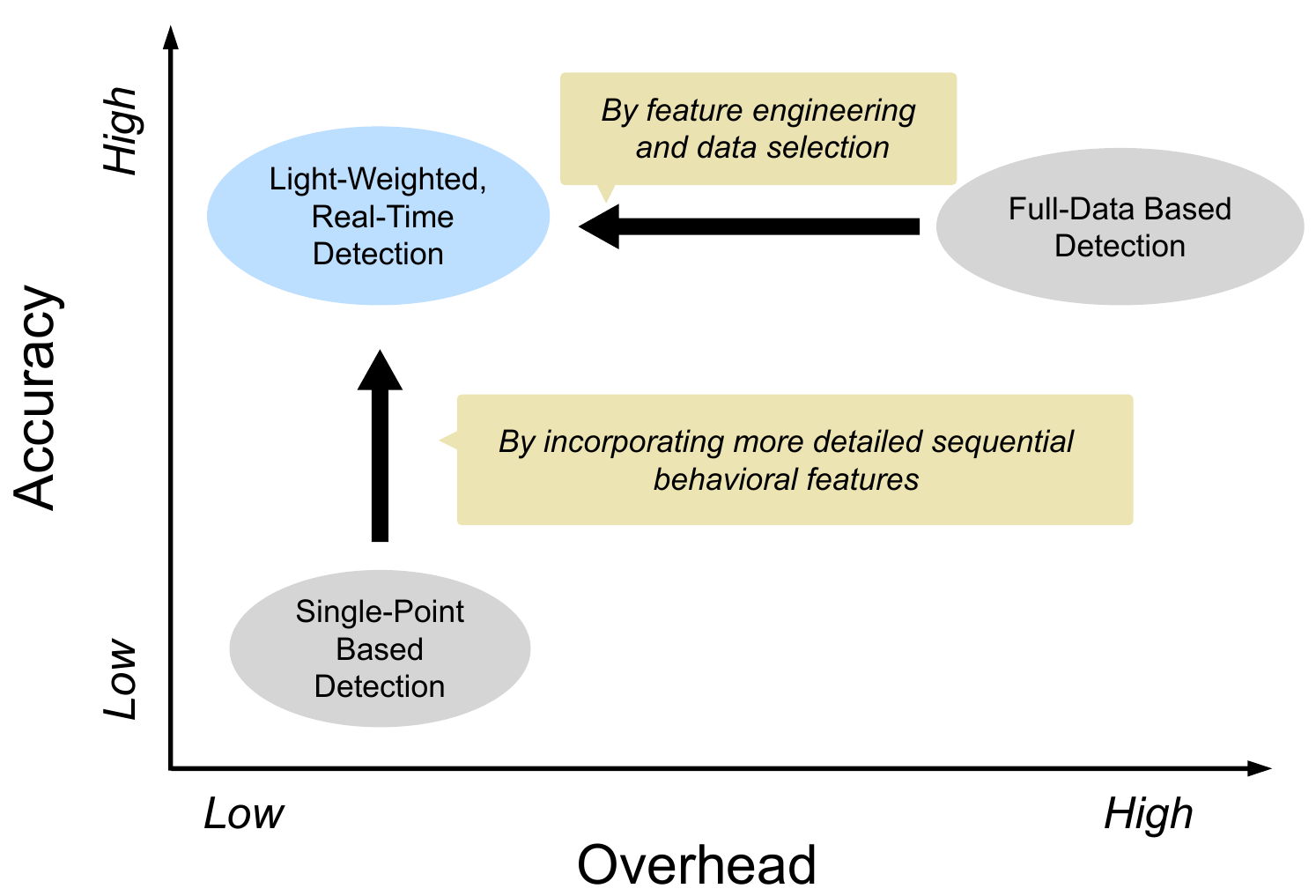}
    \caption{Trade off between two strategies}
    \label{fig:trade}
\end{figure}

To address the abovementioned challenges, we design PARIS, the first practical, adaptive trace-fetching, real-time malicious behavior detection system.
Specifically, We design several methods for feature selection, such as graph-based API selection, API association analysis, call stack selection, and loop compression, to filter out irrelevant APIs and call stacks during collection.
Moreover, based on ETW, we build an efficient, selective call stack parsing module for data collection.
Therefore, PARIS can efficiently collect and process API call stacks and perform stable, accurate, real-time behavior detection with low overhead.
By analyzing the API call stacks, we aim to identify attack behaviors rather than merely detecting the malware.
To concentrate on malicious behaviors, we devise a clustering-based training set cleaning mechanism to eliminate non-malicious behavior data (noise and usual background activities) from the training set.
Finally, all the above-mentioned methods are based on basic observation and common sense in API analysis, requiring no expert knowledge about specific API functions during detection.
Our final goal is to implement an adaptive, lightweight and low overhead real-time detection system as shown in Fig.\ref{fig:trade}.

We deploy PARIS in a real-world environment and conduct experiments with benign and malicious datasets. 
The results further show that PARIS retains only the equivalent of 1.12\% of the original data size and can run stably on the client for a long time with an average resource overhead of 32MB memory usage and 4.79\% CPU usage.
In addition, PARIS transmits at an average network bandwidth of 0.77kb/s, achieving a detection accuracy of 93.6\%, which is comparable to offline methods.


All in all, we make the following contributions in this paper:
\begin{list}{\labelitemi}{\leftmargin=1.5em}

\item We design PARIS, a lightweight real-time malicious behavior detector.
PARIS can dynamically monitor all system-level API calls (API calls in all DLL files under \texttt{C:\textbackslash Windows}) with low overhead and detect threats in real-time.
By selectively processing the ETW data, we address the issue of high overhead and delay in fine-grained tracing data collection.


\item We design algorithms to analyze and select useful APIs and call stacks for behavioral detection correspondingly.
Based on the feature selection and extraction techniques in machine learning and data mining, our model does not introduce any human prior knowledge or expertise during data collection and model training, thus having less bias and stronger generality.


\item We implemented the PARIS and evaluated it in real-world environments.
The experimental results demonstrate that the data collector of PARIS can run on a standard computer with an average memory usage of 32MB, bandwidth of 0.77kb/s, CPU usage of 4.79\%, and an average detection latency of 6.84s.
Furthermore, it still achieves a high accuracy of 93.6\% to the baselines on real-world software behavior datasets, while retaining less than 2\% of the original data scale.

\end{list}

The remainder of this paper is organized as follows.
We first describe the preliminary knowledge about the harmful behaviors of APT attacks and the Event Tracing for Windows in \S\ref{PRELIMINARY}.
Then, we present a system overview in \S\ref{systemoverview}.
Next, we introduce our design and implementation of the malicious behavior semantic model and the detection model in \S\ref{training phase}, \S\ref{testing phase}, and \S\ref{implementation}, respectively.
We evaluate PARIS in \S\ref{evaluate}.
The discussion, related work, and conclusion are presented in \S\ref{Discussion}, \S\ref{RELATED}, and \S\ref{CONCLUSION}, respectively.

%% file: 2_background.tex
\section{Background}\label{PRELIMINARY}
\subsection{Malicious Behaviors in APT Attacks}

\begin{figure}[h!]
    \centering
    \includegraphics[width=0.5\textwidth]{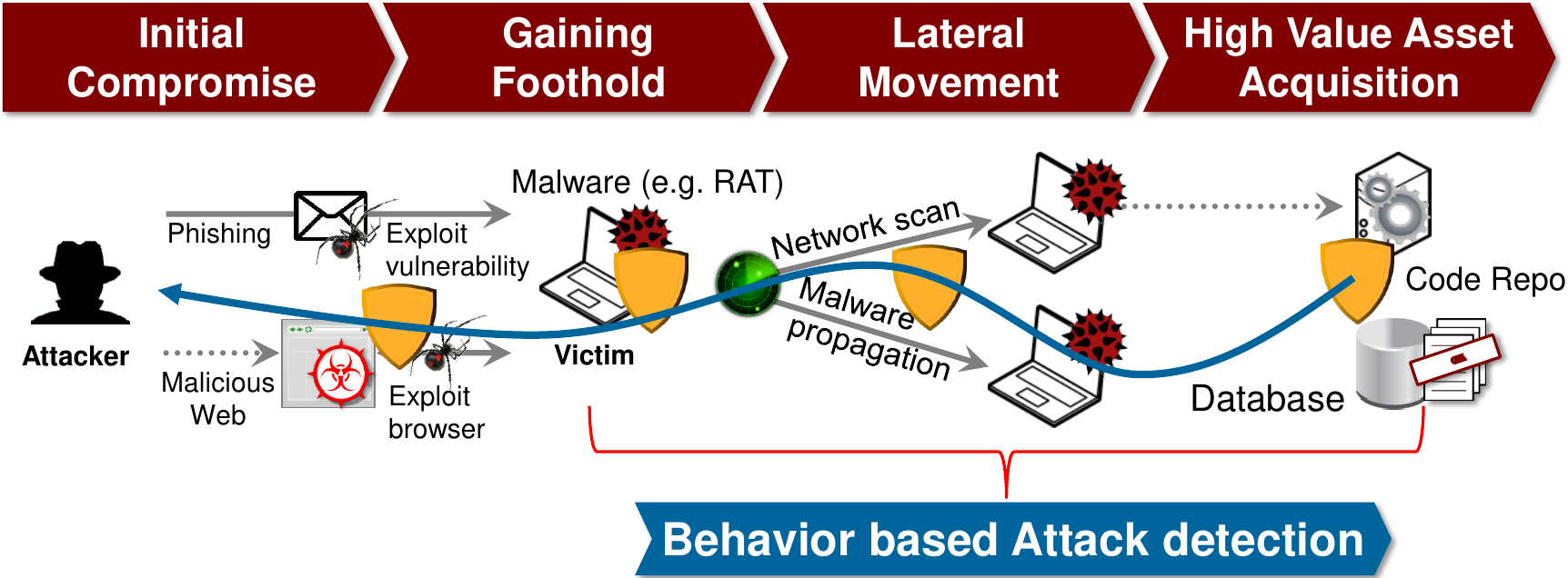}
    \caption{APT attack lifecycle.}
    \label{fig:attackflow}
\end{figure}

As Fig.~\ref{fig:attackflow} shows, the life cycle of an APT attack can be roughly divided into four stages:
(1) \textit{Preparation Stage}: prepare the attack vectors or vulnerabilities.
(2) \textit{Initial Compromise}: infect the victim hosts.
(3) \textit{Gaining Foothold}: move laterally within the network through exploits.
(4) \textit{High-Value Asset Acquisition}: identify high-value assets, and exfiltrate them.

However, these stages do not necessarily occur within a short time.
According to MITRE ATT\&CK~\cite{MITRE,TTPs}, after successfully infecting the target host, attackers often remain dormant within the system for an extended period before launching assaults.
Consequently, the dynamic detection of ongoing malicious behaviors becomes particularly critical due to the following two reasons:
(1) Given the diverse techniques employed by attackers, traditional detection methods that rely on static features (such as file hashes) or system vulnerabilities are narrow in scope and can be readily evaded by attackers.
However, once attackers gain access to a system, their behaviors are usually restricted and difficult to conceal.
For example, regardless of the technique an attacker employs for deploying their attack code, specific malicious behaviors, such as keylogging, establishing a remote shell, or communicating back with the attacker's server~\cite{xiong2020conan,yang2020ratscope}, are necessary to achieve the attack objectives.
(2) APT-related malware does not launch attacks during its dormant period.
Therefore, it doesn't show any malicious features, making it challenging to detect.
Therefore, it is more significant to focus on detecting dynamic malicious behaviors.


As described in \S\ref{introduction}, we have found that the capabilities of malware are divided into dozens of relatively independent functions, which we have named Potential Harmful Functions (PHF).
However, current APT forensics and detection systems~\cite{milajerdi2019holmes,zeng2021watson,xiong2020conan,yang2020ratscope} face challenges in accurately identifying PHFs in real-time, which are necessary for understanding the attackers' tactics and intentions to make remediation decisions.
Therefore, in this paper, we want to identify attack behaviors (PHFs) to improve the capability of APT defense.
Based on guidance from MITRE ATT\&CK~\cite{MITRE} and previous researches~\cite{whitepaper,threatpost,yang2020ratscope,}, we selected several typical PHFs that are commonly used in APT attacks, such as keylogging(T1056.001), screen stealing (T1113), remote shell (T1059, etc.), etc to verify the effectiveness of PARIS.

\subsection{ETW-based Audit Logging}
To record and analyze the fine-grained behaviors of the malware with low overhead, we adopt Event Tracing for Windows (ETW) as the data source.
ETW is a built-in log event framework based on Windows that provides detailed tracing of computer programs~\cite{etw1}.
Operating within the Windows kernel, it is optimized for high performance, boasting two key benefits: non-intrusive modifications and minimal system load.
Therefore, it is widely used in existing attack investigation work~\cite{ahmed2021peeler} and commercial enterprises such as Docker, AWS, and MS SQL Server.

\begin{figure}[h!]
\vspace{-0.1in}
    \centering
    \includegraphics[width=0.45\textwidth]{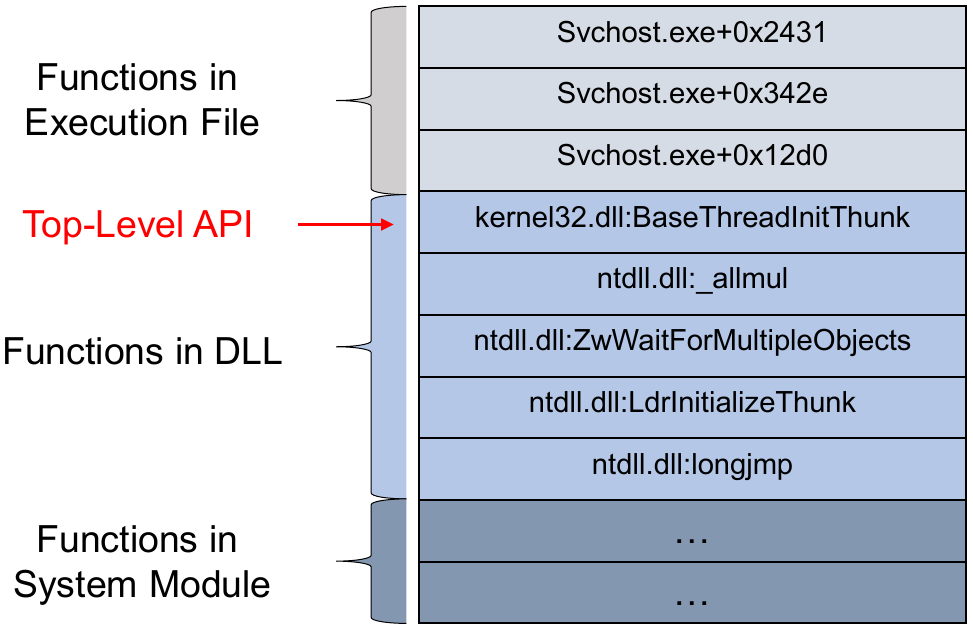}
    \caption{ A parsed full call stack example in ETW event.}
    \label{fig:CSexample}
\end{figure}

The Application Programming Interface (API) call stack (CS) is one of the most crucial data sources for dynamic detection.
As shown in Fig.~\ref{fig:CSexample}, a call stack, from top to bottom, usually starts with functions defined in user applications, followed by API functions in the system libraries, and ends with system calls.
In this paper, we define the top-level API as the first system library API function invoked by the application.
Since the attackers can deliberately change the name of the user-defined API to avoid detection, our system excludes these APIs in the collection stage.

In existing work, ETW is used to collect coarse-grained events related to operating system objects (entities), such as process creation, file read and write, and memory allocation~\cite{9152771,wei2021deephunter,wang2020you}.
However, using coarse-grained data significantly diminishes behavior detection accuracy.
In the case of fine-grained data, such as the API call stack and system calls, the ETW native parser also encounters challenges in both data parsing performance and data quality~\cite{kwon2018mci, kolbitsch2009effective}.

To address these challenges, we design and implement a parsing module for ETW raw binary data, which achieves efficient parsing and dynamic behavior restoration.
At the same time, we reduce the huge overhead brought by the system's voluminous APIs and call stacks, realizing adaptive API and call stack selection.

\subsection{Motivation Example}
Consider the following attack scenario, where the attacker delivers a Word file carrying a link to a malicious file via email, and induces the user to trust and visit the malicious link to download the payload file, which is then unzipped and run.
The victim directly double-clicked the executable attachment, and the malicious attachment was executed, first in the foreground to open a Word document to confuse the user and, at the same time inject itself into the Internet Explorer process to bypass the firewall, and connected to the attacker's server \textit{ip:port}.
In addition, the attacker launched his own \textit{malware.exe} and connected to another remote server.
Then, the attacker started to perform malicious behaviors to collect private system data and send it.
\begin{figure}[h!]
    \centering    \includegraphics[width=0.45\textwidth]{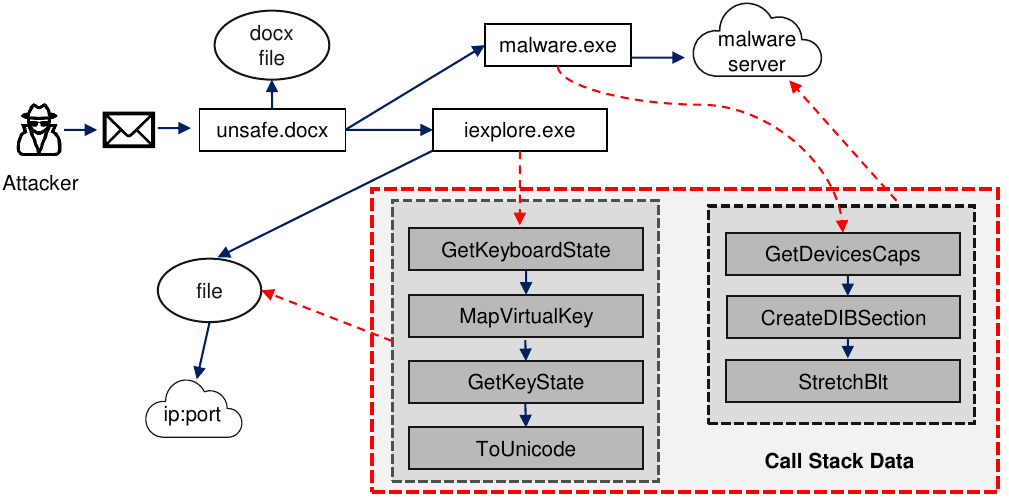}
    \caption{Motivation Example. (The red dotted box shows the process call stack information)}
    \label{fig:motivation}
\vspace{-0.1in}
\end{figure}

The specific attack process is described in Fig. \ref{fig:motivation}, where the red dotted line marks the part of the process behavior that cannot be tracked and captured by the system's high-level events.
For example, the attacker uses the \textit{iexplore.exe} process to interact with the device through the available APIs provided by the operating system to capture the user's keyboard input, which is subsequently written to disk and sent to the attacker's server.
The \textit{malware.exe}, on the other hand, uses the system-provided API to perform screen capture behavior while transferring data in real-time to the attacker's server and leaving no trace on the local disk.

Existing auditing systems \cite{liu2018towards,milajerdi2019holmes,han2020unicorn} are only able to find processes and files related to the attacker, and it is difficult for security practitioners to determine whether the system has received an attack based only on this process and file information. 
If we disregard call stack information and rely only on high-level system events for attack determination, we cannot detect this attack, nor can we understand the attacker's tactics and intent and respond accordingly.

To accurately detect malicious behaviors and to complement existing security auditing systems, we propose PARIS, a real-time and non-intrusive behavioral detection system, which uses process call stack information to analyze the fine-grained semantic behaviors of processes.
Experimental results show that PARIS can accurately identify fine-grained semantic behaviors of processes and introduces only acceptable system load.

%% file: 3_method.tex
\section{System Overview}\label{systemoverview}
In this section, we present the assumptions and threat model underlying PARIS and outline the structure of our system, including behavior model training and real-time detection.

\subsection{Threat Model}\label{threadmodel}
We assume that the underlying operating system and audit system are part of the Trusted Computing Base (TCB), which means that ETW will not be tampered with or disabled by attackers.
Such assumption is shared among studies related to system auditing and intrusion detection~\cite{zeng2021watson,zengy2022shadewatcher,hassan2019nodoze,hassan2020omegalog,yang2020ratscope}.

Our real-time detection refers to blocking attacks by identifying malicious behavior within a complete attack window, and notifying AV/EDR to respond, thereby enabling real-time response and interruption of the attack process.


\subsection{Framework}
Our system consists of two parts: process behavior modeling and process behavior detector.
The first part aims to capture the features of malicious behaviors in real-time with low overhead.
The goal of the detector is to detect malicious behaviors while saving running time and system resources as much as possible.

\textbf{Adaptive Trace Fetching.}
In this phase, we propose an ETW-based collector to collect API call stacks.
Through events filtering, caching, and automated analysis of API call stacks, we achieve adaptive selection and efficient parsing, realizing lightweight and real-time data collection.
Note that our adaptive trace-fetching model is not only applicable to behavioral detection in this paper but also facilitates attack analysis and forensics in general.

\textbf{Behavior Identification.}
In the adaptive trace-fetching phase, we get concise and representative call stack data, which is used in feature extraction and behavior detection.
We implement a real-time classification detector based on \textit{Random Forest}.

\begin{figure*}[h!]
\vspace{-0.1in}
    \centering
    \includegraphics[width=0.9\textwidth]{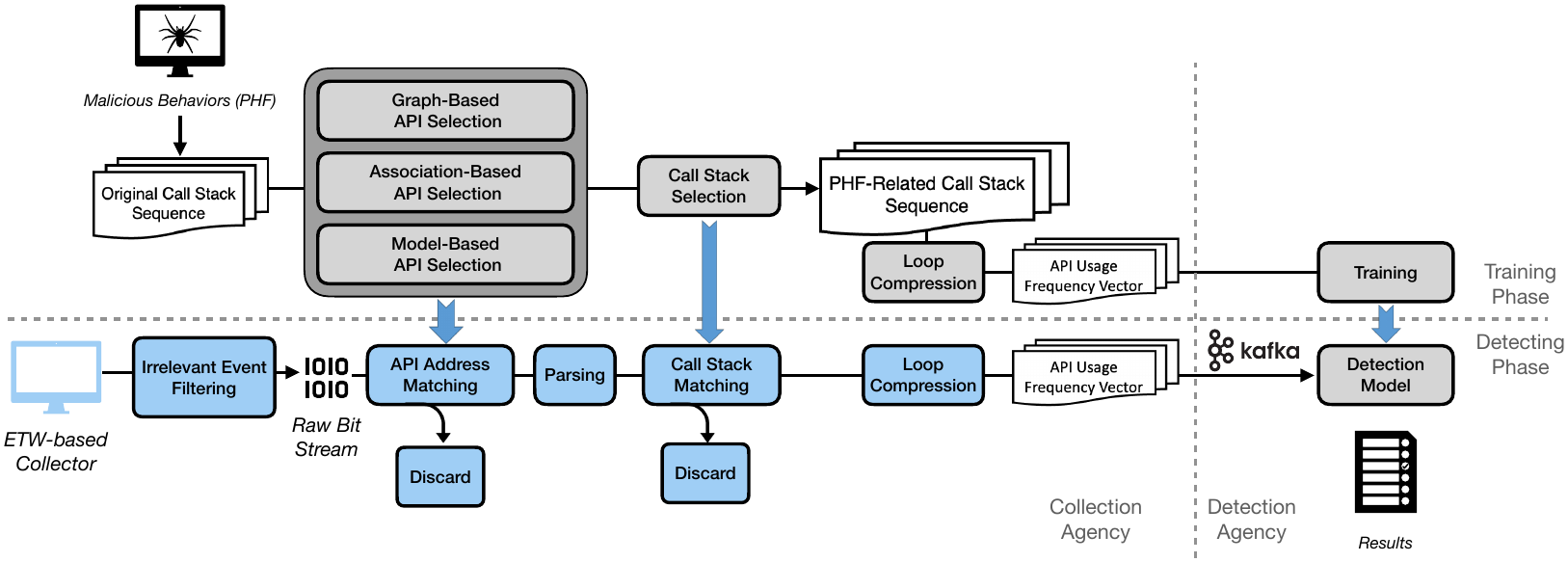}
    \caption{The structure of PARIS, including the behavior modeling part (training phase) and real-time detecting part (detecting phase). (The blocks in blue represent the data processing pipeline before detection.)}
    \label{fig:framework}
\vspace{-0.1in}
\end{figure*}

As shown in Fig.~\ref{fig:framework}, for the malicious behavior modeling part, there are malware data collection and parsing, graph-based API selection, association-based API selection, call stack selection, loop compression, and feature embedding.
The details of this part are demonstrated in~\S\ref{training phase}.
The real-time behavior detector is constructed by the following parts: software data collection, API filtering, call stack filtering, loop compression, feature embedding, and detecting.
More details can be found in~\S\ref{testing phase}.

\section{Malicious Behavior Modeling}\label{training phase}
\begin{figure}[h!]
    \centering
    \includegraphics[width=0.45\textwidth]{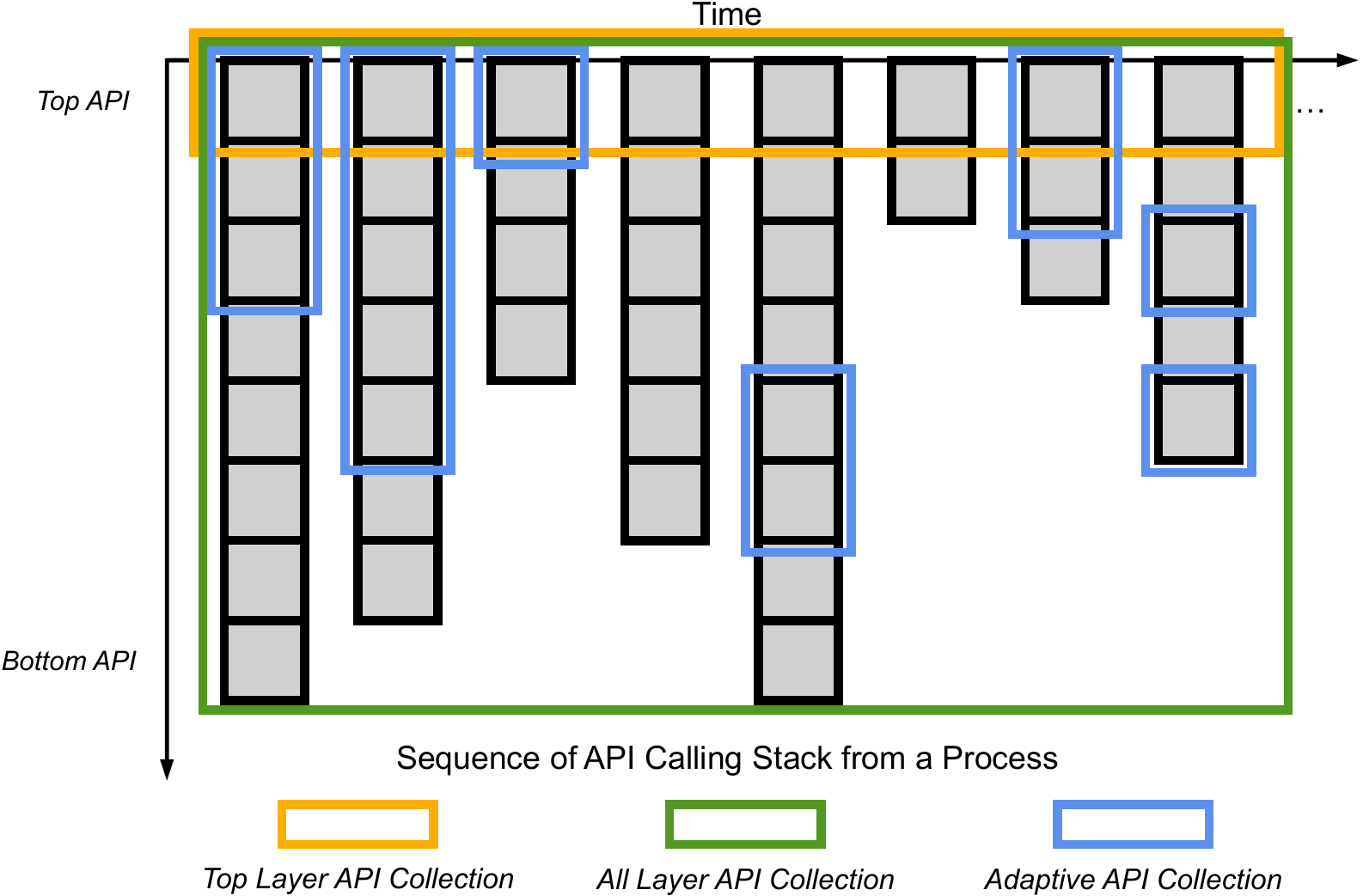}
    \caption{The basic idea of adaptive data collection.}
    \label{fig:adaptive_collection}
\end{figure}

An important contribution of our work is the adaptive trace fetching in a real-time detecting system, which is shown in Fig.~\ref{fig:adaptive_collection}.
Typically, all system logs (in this paper, API call stacks) are fetched to identify the malware features and get the best detection model~\cite{yang2020ratscope}.
However, it causes a lot of memory and CPU time to parse, recover, and cache the data.
In real-world APT detection systems, it is common to grab the top-level API to control the overhead of the collector~\cite{xiong2020conan}.
However, this straightforward selection strategy brings a significant drop in detection accuracy.
In this paper, we aim to automatically learn the critical APIs and call stacks for malicious behavior detection and skip the irrelevant ones.
We avoid excessive overhead by finding the most representative data with our adaptive trace-fetching model. 

We illustrate our system in the following sections.
For each section, we design some methods or use some techniques based on observations or experimental verifications.
We first demonstrate the basic observations we rely on, and then the description of our methods follows. We will show the effectiveness of our design in \S\ref{evaluate}.
\begin{figure}[h!]
    \centering
    \includegraphics[width=0.4\textwidth]{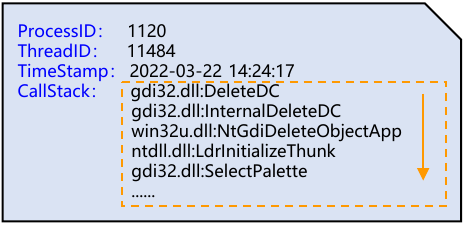}
    \caption{An example of ETW event generated by our audit logging system.}
    \label{fig:call stack Data}
\vspace{-0.1in}
\end{figure}

\begin{algorithm}
    \caption{Call Stack API Parsing Algorithm}\label{alg:callstack-api}
    \SetCommentSty{small}
\SetCommentSty{small}
\KwData{Call Stack: $CS_o$}
\KwResult{Call Stack data after dynamic parsing: $CS_n$}
$PID,TID,TimeStamp,Address_{Stack} \gets CS_o$\;
$CallStack \gets stack[\ ];$
$L=0$\;
$UseAPI_{cache} \gets [\ ];$
$UselessAPI_{cache} \gets []$\;
\While{$L < $ len($CS_o$)}{
    $L=L+1$\;
    $Address_L \gets CS_o[L]$\;
    \If{$Address_L \ in\  UselessAPI_{cache}$}{
        $continue$\;
    }
    \eIf{$Address_L \ in\ UseAPI_{cache}$}{
        $API \gets UselessAPI_{cache}$
    }{
    $DLL_L \gets DLLSearch(Address_L)$\;
    $API_L \gets APINameSearch(Address_L)$\;
    $API \gets DLL_L + API_L$\;
    }
    $API_{Index} = APISelection(API)$\;
    \tcp{Determines if the API is irrelevant, returns -1 as irrelevant}
    \eIf{$API_{Index}!=-1$ AND $API_{Index} != Pre_{APIINDEX}$}{
        $CallStack.pushback(API)$\;
        $Pre_{APIINDEX} \gets API$\;
        $UseAPI_{cache} \gets Address_L$
    }{
    $UselessAPI_{cache} \gets Address_L$
    }
}
\tcp{Determine if this is an irrelevant call stack, irrelevant is not returned}
\If{$isUselessCallstack(CallStack)$}{$return$\;}
$return \ CallStack$\;
\end{algorithm}

\subsection{PHF-Irrelevant events reduction}
ETW provides over a thousand types of events.
Most of them are specific to certain applications, such as Internet Explorer and Microsoft Word, while our model depends on general events (e.g., system calls and call stacks) to represent PHF behaviors~\cite{yang2020ratscope,zhu2021general}.
Therefore, we reduce the audit log by filtering out events unrelated to malware behavior and focusing on the remaining system call and call stack events.

The fields we reserve are processID, threadID, timestamp, and call stack.
The format of event data is shown in Fig.~\ref{fig:call stack Data}.
By filtering out the PHF-irrelevant ETW events, we remove a lot of redundant data and save many system resources.

\subsection{API Selection}
There are primarily two steps for selecting APIs from the call stacks: removing \textit{trivial functional API} and removing \textit{semantically redundant API}.

The first step is to remove APIs that are uncorrelated to specific call stacks.
As many APIs are needed for trivial functions, they commonly appear in most call stacks.
Because they are widely called in nearly every call stack, these APIs can not provide any helpful information for distinguishing a malicious process.
Take \texttt{ntdll.dll:LdrInitializeThunk}, the entry point of \texttt{ntdll.dll}, as an example.
Whenever the system wants to call other APIs or some APIs in unloaded DLL files, it usually needs to call it first~\cite{noauthor_msdn_nodate}, while it has nothing to do with the specific behaviors.


To filter out these unimportant APIs, the significance of different API functions needs to be evaluated. 
An intuitive assumption is that the more frequently an API appears in different call stacks, the more ordinary its function is.
However, some call stacks may have similar functions and look slightly different, causing the higher frequency of corresponding APIs.
Given this, the frequency-based importance is unsuitable.
Alternatively, we put forward a graph-based API importance evaluation method.
We first build a graph based on the set of all call stacks.
The vertices are the API functions in this set, and the edge between two nodes is determined by whether the two API functions are called consecutively in a single call stack.
If so, then we build an edge between these two API functions, which shows that they have some functional relationship.
The advantage of this graph-based method is that even if there are many similar call stacks occurring in the call stack set, the frequency will not change the relationship between the majority of API functions.

After that, we evaluate the importance of every API function (vertex in the graph) by the following equation
\begin{equation}
\small
    I_v = Degree(v)/(N-1)
\end{equation}
where $N$ is the total number of vertices and $Degree(v)$ is the degree of vertex $v$.
From the formula, it can be inferred that the greater $I_v$ is, the more likely the API is positioned at the center of the "API calling community," suggesting that the API tends to undertake trivial functions.

Here we show some trivial API functions learned by our method and a brief introduction to their semantics according to ~\cite{noauthor_msdn_nodate} in Table~\ref{fig:api_importance_low} (A more complete list can be found in Table~\ref{tab:APIimportance} in the Appendix).
From their semantics, we know that the graph-based importance can capture those trivial API functions that have nothing to do with the process behavior.
We set a threshold and then filter out all trivial API functions from the training set.
During the detection stage, we would also ignore these trivial APIs to save more machine resources and reduce the overhead.

\begin{table*}[h!]
 \caption{Description of API functions with minimal correlation to specific call stacks and behaviors.}
    \label{fig:api_importance_low}
\centering
\begin{tabular}{@{}lcl@{}}
\toprule
API Name                                & $I_{v}$     & Description                                                    \\ \midrule
kernel32.dll: BaseThreadInitThun            & 0.636 & Calls the thread’s entry point.                               \\
ntdll.dll: LdrInitializeThun                & 0.658 & Starts threads of user mode.                                  \\
ntdll.dll:RtlGetAppContaine rNamedObjectPat & 0.642 & Retrieves the named object path for the app container.        \\
ntdll.dll: KiUserCallbackDispatcher         & 0.655 & Passes message information to the specified window procedure. \\
wow64cpu.dll:BTCpuSimu                      & 0.655 & Support for running x86 programs on x64.                      \\ \bottomrule
\end{tabular}
\end{table*}

\subsection{API Association Analysis}\label{api-association}
As mentioned before, the second step for the API selection is to remove semantically redundancy: the association/dependency between different API functions.
By association, we refer to the fact that some APIs may show up simultaneously in a call stack~\cite{ye2007imds}.
The reason is the dependency relationship among different API functions during the calling procedure, e.g., Calling $API_A$ is the prerequisite for calling $API_B$ or $API_A$ function has a tail jump to $API_B$~\cite{cono2020designing}.

We use a real-world example to show this association.
There are two system API functions from Windows: \texttt{GetMessageW} from \texttt{user32.dll} and \texttt{NtUserGetMessage} from \texttt{win32u.dll}.
They almost always show up together because their functions are very similar: retrieving a message from the message queue of the calling thread.
In fact, what \texttt{GetMessageW} does when called by a thread is just to call \texttt{NtUserGetMessage}, and \texttt{NtUserGetMessage} would find the corresponding system call.
Therefore, if we have already got \texttt{GetMessageW}, which is always followed by \texttt{NtUserGetMessage} from a call stack, we usually don't have to waste any computing resources to parse, process, or save the second one.

To analyze the association relationship between $API_A$ and $API_B$, we define three values: Support $S(A)$, $S(A, B)$, Confidence $C(A\rightarrow B)$, and Lift $L(A\rightarrow B)$ as
\begin{equation}
    S(A)=Freq(A)/T
\end{equation}
\begin{equation}
    S(A, B)=T(A, B)/T
\end{equation}
\begin{equation}
    C(A\rightarrow B)=S(A, B)/S(A)
\end{equation}
\begin{equation}
    L(A\rightarrow B)=S(A, B)/S(A)S(B)
\end{equation}
where $Freq(A)$ represents the number of records in the dataset that contains A, $T(A, B)$ is the number of call stacks that contain both $API_A$ and $API_B$, $T$ is the total number of call stacks.
From the equations, we can know that
\begin{enumerate}
    \item If $API_A$ and $API_B$ occur simultaneously in a call stack more, $S(A, B)$ would be higher;
    \item If $API_B$ occurs more under the condition that $API_A$ occurs, $C(A\rightarrow B)$ would be closer to 1;
    \item If $API_B$ is more likely to occur with $API_A$ than with other API, $L(A\rightarrow B)$ would be higher.
\end{enumerate}
If $API_A$ and $API_B$ have a high value of $L(A\rightarrow B)$, that means $API_B$ is highly determined by $API_A$.
Therefore, we are confident to remove $API_B$ or $API_A$ from the API collection set if we got a high $L(A\rightarrow B)$ and a high $L(B\rightarrow A)$ because we know that the information brought by them is almost identical.

In order to capture the dependency without any prior domain knowledge, we use the Apriori algorithm~\cite{agrawal1996fast} to find the association rules among the system call stack logs.
Briefly, Apriori gives an efficient way to find the most frequently occurring set without specifying the set's size.
The main idea is the subsets of a frequently-occurring set are also frequently-occurring. More details about this algorithm can be found in many tutorials and efficient implementations~\cite{bodon2003fast,borgelt2002induction,hegland2007apriori}.
We showed the result of our API association analysis in the Appendix (See Table \ref{tab:APIassociation} for details).

\subsection{Select API from Detection Model}\label{sec:selectfrommodel}
Many machine-learning based models assign importance to each feature during the training of classification or regression.
These importances, which may have different conceptual meanings, show the contribution of each feature to building a reliable and effective model.
For the linear models, the importance of a feature is the absolute value of the corresponding coefficient.
While for the tree-based estimators and the ensemble models based on forests, the Gini importance can serve as the feature importance.
To get the importance of different APIs, we perform frequency statistics on the logs and calculate the API-frequency vectors, where each API corresponds to a feature dimension in the vector.
We then feed these vectors to the machine learning models to evaluate the feature importance.
Finally, we set a threshold to filter out the APIs whose importance is below it.
In other words, only those APIs which have high importance are collected by the data collector.
We use various models for feature selection.
Their performance, as well as the selection of threshold, are discussed in \S\ref{sec:ablation}. 
We show the main flow of the real-time API selection and filtering process in Alg.\ref{alg:callstack-api}.

\subsection{Call Stack Selection}\label{sec: call stack selection}
For the call stack selection module, it consists of two parts.
First, we remove the \textit{unrelated call stacks} through the correlation analysis model.
We define unrelated call stacks as the call stacks that occur frequently in both benign and malicious software or occur in different PHFs simultaneously.
We are inspired by the idea of \textit{inverse document frequency (IDF)}, which is widely used in many NLP-based malware detection works~\cite{tran2017nlp}.
The basic idea of IDF is to reduce the significance of highly common words in the collection while enhancing the importance of seldom-occurring words.
Thus, we first calculate the distribution for each call stack and create formal definitions to quantify unrelated call stacks using the benign software labels and the PHF labels in the training set.
We define the \textit{behavior correlation index (BCI)} and \textit{malicious behavior correlation index (MBCI)}.
\begin{equation}
\vspace{-0.05in}
    BCI_{sc} = \sum_{class} \frac{n_{class}^{sc}}{N_{class}} \times \log \frac{n_{class}^{sc}}{N_{class}}
\vspace{-0.02in}
\end{equation}
\begin{equation}
    MBCI_{sc} = \sum_{PHF} \frac{n_{PHF}^{sc}}{N_{PHF}} \times \log \frac{n_{PHF}^{sc}}{N_{PHF}}
\end{equation}
where, $N$ is the total number of log samples, $N_{PHF}$ is the number of malicious behavior (PHF) samples.
$n_{PHF}^{sc}$ is the number of samples that contains call stack $sc$ among all samples of this PHF.
The calculation of BCI and MBCI bear certain similarities to the entropy in information theory. 
According to the equations, a higher BCI or MBCI for a call stack indicates a more uniform distribution across various PHFs and software, suggesting that the call stack is less likely to be associated with any specific malicious behavior.

Secondly, based on our observation, it is very common to find many loops in the call stack sequences.
For example, a device driver may keep pooling some specific ports to check the status of the device.
Or, a process may call \texttt{KernelBase.dll:SleepEx} repeatedly when waiting for the new commands from the attackers.
These repeated loops may cause a huge amount of redundancy, bringing nothing useful for us to understand the behavior of the process.
We develop a loop-compression algorithm Alg.\ref{alg:loop-detecting} to detect the duplication of subsequences and compress the replications into a single subsequence. We briefly describe the call stack selection process in Alg.\ref{alg:callstacks}.


\SetKwComment{Comment}{/* }{ */}

\begin{algorithm}
\caption{The Loop-Compression Algorithm $LCA$}\label{alg:loop-detecting}
\KwData{Old call stack sequence: $CSseq_o$}
\KwResult{New, compressed call stack sequence: $CSseq_n$}
$CSseq_n \gets [\ ]$\;
$lastAppearanceIndex \gets dict\{\}$\;
$i \gets 0$\;
\While{$i < $ len($CSseq_o$)}{
$cs \gets CSseq_o[i]$\;
\eIf{$cs \in lastAppearanceIndex$}{
    $i_l \gets lastAppearanceIndex[cs]$\;
    $lastAppearanceIndex[cs] \gets i$\;
    Compare $CSseq_o[i_l:i]$ and $CSseq_o[i:2i-i_l]$\;
    \eIf{Matched}{
        $i \gets 2i-i_l$\;
    }
    {
        $k \gets$ the index of first unmatched event.\;
        $CSseq_n \gets CSseq_n||CSseq_o[i:k]$\;
        $i \gets k$\;
    }
}
{
$lastAppearanceIndex[cs] \gets i$\;
add $cs$ to $CSseq_n$\;
$i \gets i + 1$;
}
}
\eIf{len($CSseq_n$) $<$ len($CSseq_o$)}{
    \Return $LCA(CSseq_n)$\;
}
{
    \Return $CSseq_n$
}
\end{algorithm}

\section{Detection Model}\label{testing phase}
\subsection{Feature Embedding}
We get the more succinct and representative sequences of refined call stacks from \S\ref{training phase}.
Now we need to find out which sequence shows the malicious behavior.
Before considering any classification model in machine learning and pattern recognition, we need to extract feature vectors from the refined call stack sequences.

In this work, we use the frequency of API usage for feature extraction following the previous work~\cite{tian2010differentiating,sami2010malware} for the following reasons:
\textbf{Lightweight}: without any matrix computation, the frequency of API usage requires much fewer system resources compared with all deep-learning-based models.
\textbf{Agile}: generating a frequency vector is much quicker than other models.
Therefore, it is suitable for real-time analysis and detection.
\textbf{Accurate}: one drawback of the frequency of API usage is its incapability of analyzing the order information of the sequence.
However, it would not hurt the accuracy much when we took low-level information into consideration.
Note that we monitored all DLL files under \texttt{C:\textbackslash \textbackslash Windows} folder.
After removing the noise, it is very hard to achieve new behavior without importing new DLL files and API functions.
In other words, it is hard to use exactly the same system-level call stacks, just in a different order, to achieve different behaviors if we monitor all system-level API calling functions.

\begin{algorithm}
    \caption{Algorithms for analyzing call stacks}\label{alg:callstacks}
    \SetCommentSty{small}
\KwData{Sequence of call stacks for all processes: $CSseqs_o$}
\KwResult{Feature vector for all processes: $ProcessFeature$}
$PID,TID,CS_{Sequences} \gets CSseqs_o$\;
$ProcessFeature \gets dict\{\}$\;
\tcp{After removing irrelevant call stacks}
  \For{each process's call stack Sequences}{
    $ CS_{R} \gets RemoveDuplicateCS(CSseqs_o)$\;
    $ CS_{RL} \gets LoopCompression(CSseqs_{D})$\;
    $ Feature_{PID} \gets FrequencyVector(CS_{RL})$\;
    $ ProcessFeature[PID] \gets Feature_{PID}$\;
  }  
$return \ ProcessFeature$\;
\end{algorithm}

\subsection{Classification Model}
After getting the embedded vectors, many machine learning models can be used to classify those feature vectors, including Support Vector Machines (SVM), Neural Network models, Ensemble methods, and so on.
We perform an experiment to compare different machine learning models to choose the best detection model.
After trying many commonly used machine learning classification models, we selected Random Forest as our classifier for its high accuracy.
The experimental details can be found in \S\ref{sec:ablation}.
What's more, the measurement of the feature importance given by Random Forest allows us to choose important APIs, further narrowing down the scope of API collection and achieving a lower collection load.

%% file: 4_implementation.tex
\section{Implementation}\label{implementation}

\subsection{Data Collector}
There are many ways to collect running traces of the process dynamically, such as hooking, sandboxes, taint tracking, traceability diagrams, and so on.
Due to their significant computational requirements, these methods are not appropriate for implementation in a low-cost, real-time detection system over an extended period.
Moreover, the protection of the Windows system at the kernel level has been continuously strengthened, and the kernel protection Patch (KPP) developed by Windows has also increased the difficulty of obtaining data by the above method.
Therefore, most of these methods make intrusive modifications to the system, and the program behavior traces they collected are relatively coarse-grained and not detailed enough to achieve accurate detection of malicious behavior, and the generalization ability of the detection model is also weak.

In order to solve the problem of fine-grained, real-time semantic restoration, we implemented a series of efficient data parsing, redundant data removal, and transmission modules based on native ETW on Windows.

\subsubsection{Efficient ETW data parsing and semantic restoration}
Coarse-grained log data is too ambiguous to reflect the behavior of a process.
Therefore, we used the real-time data collection and analysis module based on ETW to provide fine-grained behavioral traces of the process.

\begin{figure}
\vspace{-0.1in}
    \centering
    \includegraphics[width=0.45\textwidth]{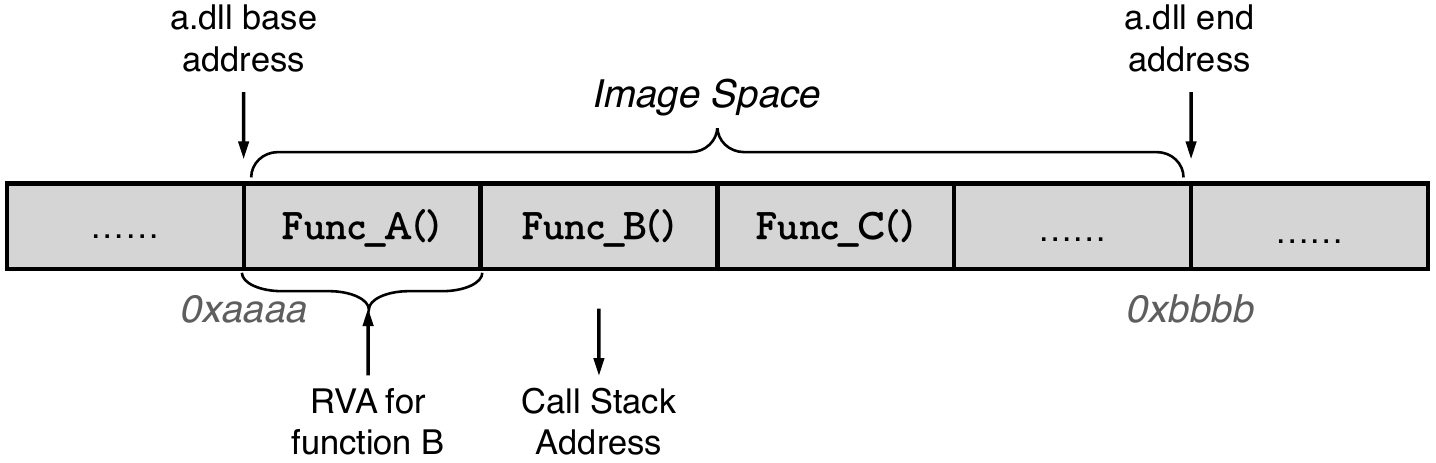}
    \caption{Address space of loaded DLL and API offset.}
    \label{fig:DLL}
\vspace{-0.1in}
\end{figure}

The call stack walking module in ETW provides the addresses of the entire dispatch stack~\cite{cswalk}, which needs to be parsed into the corresponding API function names.
Our program parses the multi-layer API from the application address space to the kernel space. 
In order to parse the API name corresponding to the address in the call stack, it is necessary to obtain the DLL files loaded into the address space and the $base\ address$ of the DLL.
We designed and implemented the \textit{callstackTraceGenerate} program to obtain the DLL RVA.
By using this, we can get the RVA--API of the exported function in the DLL, and use the $base\ address + rva$ to get the corresponding API name.
Referring to Fig.~\ref{fig:DLL}, the address space of Func\_B is $base\ address+ rav_b$.
When the address of an API function in the call stack is in this range, it is resolved to Func\_B.
\par
Our collector maintains a map, the key is the DLL name, and the value is the mapping of the RVA--API.
In order to improve the efficiency of query and insertion, the mapping of RVA--API also needs to be stored.
The key is the RVA start address and RVA end address of the Function, and the value is the API name.

By using operator overloading, calling the above addressing method can obtain the corresponding API name according to the address in the call stack.
Then, we store the mapping relationship between the address and the corresponding API semantic information in a specific cache structure, so that it can be obtained directly without parsing in the next lookup, thereby reducing the overhead on our system.

\subsubsection{Redundant data removal}
\begin{figure*}[h!]
\vspace{-0.1in}
    \centering
    \includegraphics[width=0.85\linewidth]{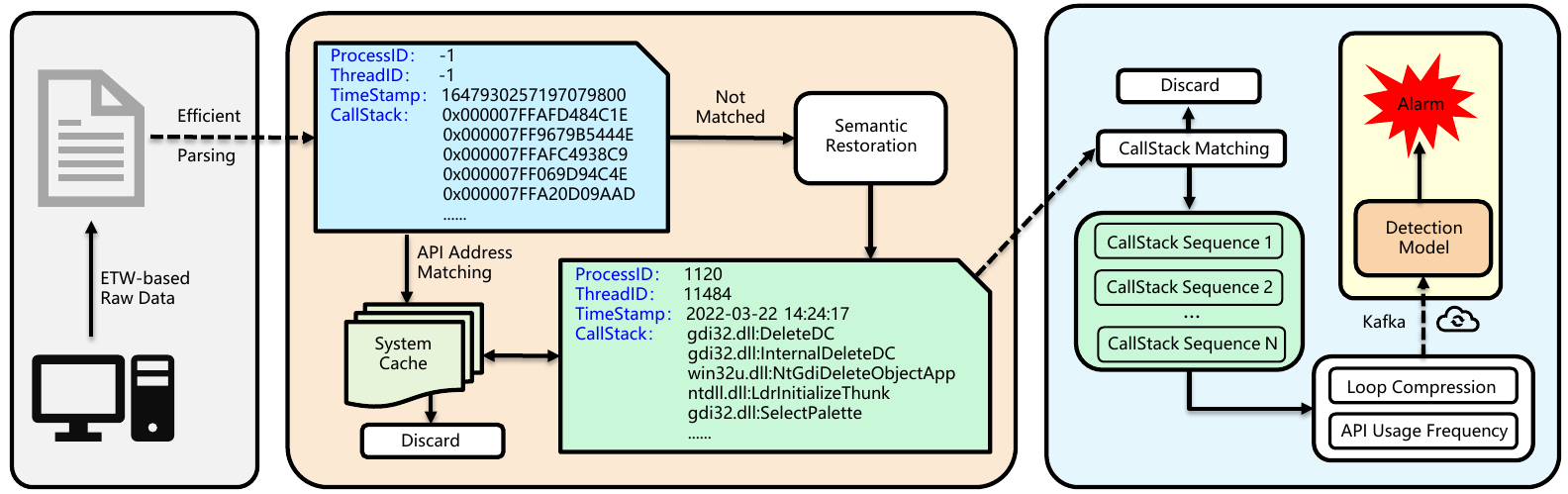}
    \caption{Practical implementation of our log auditing and detection system.}
    \label{fig:data_pipeline}
\vspace{-0.1in}
\end{figure*}

One of the performance bottlenecks when collecting call stack traces is the parsing of the API functions.
Although we did not use the native event parser provided by Windows but reimplemented and optimized the parsing module to reduce the overhead, we still wanted to further reduce the system overhead without affecting the final detection analysis. 
Therefore, we will prioritize comparing memory addresses to filter duplicate APIs. In addition, as shown in Fig. \ref{fig:data_pipeline}, we design a dynamic caching module to filter out trivial functional APIs and semantically redundant APIs in advance, rather than consuming CPU for parsing. At the same time, we will also load the Call Stack Matching module to filter data between stacks, avoid memory accumulation, and minimize overhead as much as possible.

\subsubsection{Compressed transmission of data}
Although the data is efficiently parsed and filtered, we hope to further reduce the data size for better real-time performance.
Thanks to the simplicity of frequency vectors, it is easy to generate the feature vectors locally and send them to the detection server directly.
We send compressed frequency vectors to the detection server directly instead of sending the sequence data and running a complicated classification model to meet the real-time requirement.

\subsection{Behavior Detection Module}\label{detection_implementing}
Our implementation of feature engineering (API/call stack selection, etc), model training, and detector include 6.5+KLoC of Python.
For the machine learning classification models, we use the popular Scikit-learn~\cite{scikit-learn} library, which offers a lot of machine learning models.
We introduce the selection of models and parameters through experimental results in \ref{sec:ablation}.
We will release our code (except our ETW-based data collector) after getting accepted.

%% file: 5_experiment.tex
\section{Evaluation}\label{evaluate}
In this section, we evaluated our system in terms of detection accuracy, response time, and computational overhead.
Specifically, we evaluate PARIS by answering the following questions:
\vspace{0pt}
\begin{list}{\labelitemi}{\leftmargin=1.5em}
 \setlength{\topmargin}{0pt}
 \setlength{\itemsep}{0em}
 \setlength{\parskip}{0pt}
 \setlength{\parsep}{0pt}
    \item RQ1: How accurate is PARIS in detecting malicious behaviors? (\S\ref{balance}, \S\ref{accuracy1})
    \item RQ2: How much runtime and space overhead does PARIS incur when deployed in real-world environments? (\S\ref{overhead1})
    \item RQ3: How effective is Paris in handling raw audit data and behavioral detection? (\S\ref{other})
    \item RQ4: How do different models and parameters affect the system? (\S\ref{sec:ablation})
\vspace{0pt}
\end{list}

\subsection{Methodology}
\subsubsection{Dataset}\
Based on the research \cite{whitepaper}, we found that RATs generally aggregate multiple mutually independent malicious behaviors and are used in a large number of APT attacks. Therefore, we collect a library of RAT samples from underground hacker forums \cite{hack1,hack2,hack3} to learn malicious behaviors.
At the same time, we collected data on process behavior during the operation of the corresponding benign applications.
Note that we just took the RATs or other malware that appear in real APT attacks as a collection of malicious attack functions.
But our goal is to identify attacks by restoring their behavior semantics, rather than simply detecting the RAT or malware itself. 
In the following, we describe the composition of the dataset in detail.
\newline
\indent \textbf{Malware Dataset.}
Every RAT toolkit comprises two primary components: a RAT stub and a RAT controller.
When the RAT stub runs successfully on the victim's hosts, the RAT controller can perform a series of malicious actions on it, such as keylogging, screen capturing, and establishing a remote shell, to steal sensitive information and data.
We collect 476 RAT samples in 53 categories for analysis from various sources, including SpyGate-RAT, Alusinus RAT, Dark Comet Babylon, etc.
After that, We select 21 RAT samples and deploy the RAT stub and RAT controller on two machines, respectively.
Then, we perform 105 attack behaviors individually through the controller.
At the same time, we launch the modified ETW collector to obtain the call stack data for different malicious behaviors on the victim side.

\textbf{Benign Dataset.}
For benign applications, we download and install a substantial variety of widely-used software, catering to both businesses and individual users.
First, we collect call stack data of benign applications that behave similarly to malware processes, including communication software(e.g., Outlook, Foxmail), remote accessing programs (e.g., TeamViewer, sunlogin, Xshell), text editing software (e.g., Word, Notepad++, Typora), browsers (e.g., Chrome, IE, Edge), instant message programs(e.g., Skype, Wechat, DingDing), file download tool (e.g., Google Drive, WinSCP), audio-related applications (e.g., Windows media player, music player) and so on.
We also collect long-running Windows system processes, such as cmd.exe, dllhost.exe, svchost.exe, explore.exe, and so on.
Next, we install the software on the system, simulate typical user interactions with these programs, and then collect call stack data as the benign dataset.

\subsubsection{Experimental Setups}\
We utilize the datasets mentioned above as the training and testing set, with a 70-30 split, for the evaluation.
The size of the detection window is set to 6 seconds to make sure there must be at least one successful malicious behavior during that interval.
We have not extensively engaged in parameter tuning as it is not the focal point of this paper. Instead, we have maintained consistent parameter settings for each baseline to ensure a fair comparison.
We deploy the detector and collector modules on the server and client, respectively, to evaluate the detection accuracy, detection time, and overhead, including data transmission bandwidth, memory usage, and CPU usage.
The host where we deploy our collector is based on Windows 10, with Intel i5-7500 CPU (4 cores and 3.40 GHz) and 16.0 GB Memory.

\subsubsection{Baselines}\
This is mainly because (1) they both analyze the fine-grained behaviors of the program using API call stacks; (2) they utilize full API call stacks and top-layer APIs, respectively, presenting two distinct choices between accuracy and overhead when designing API-based detection systems.

\indent 
\textbf{CONAN}~\cite{xiong2020conan}.
In order to detect PHF in real-time, CONAN only chooses the top-level API in the call stack for detection because the data volume of the full call stack is so large that real-time processing is impractical.
The idea of PHF detection in CONAN is to match the API sub-sequence with the signatures, which are defined in advance for different PHFs.
CONAN matches the collected top-layer API sequences with the signatures of PHFs using the Longest Common Subsequence (LCS) algorithm.
Thus, the detection results rely on the matching algorithm and the signatures of PHFs.
However, the definition of signatures depends on statistical analysis or expertise.
Due to the need for expertise and the poor detection performance of signature matching, in this paper, we use machine learning models to train and test based on top-layer API data to get a fair comparison.

\textbf{RATScope}~\cite{yang2020ratscope}.
RATScope uses full call stacks to detect malicious behaviors.
The system workflow consists of three stages: feature training stage, log collection stage, and detection stage.
The authors propose a program behavior model named Aggregated API Tree Record (AATR) Graph.
They use the training data to generate the AATR Graph corresponding to each PHF of each RAT.
Then, RATScope employs an optimal local graph-matching algorithm to match the generated AATR Graph with the data collected on the monitored hosts.
A successful matching means that a specific PHF behavior is detected.

\subsubsection{Metrics}\
The True Positive Rate (TPR) and False Positive Rate (FPR) are usually used to evaluate the performance of detectors.
We also use \textit{Accuracy}, \textit{Receiver operating characteristic (ROC)} Curve, and \textit{Area under the ROC Curve (AUC)} score to comprehensively balance the true positive rate and false positive rate.
We use the One-vs-the-Rest (OvR) multiclass strategy when evaluating the ROC curve in the behavior classification problem.

Another highlight of our system is its ability to perform real-time detection with a much lower burden to the system.
In order to compare the load occupancy of the data acquisition and processing modules, we deployed the log collection, parsing, and sending modules of three baselines on the actual machine, then tested the memory and CPU usage, as well as the data transmission bandwidth to show that our system can meet the real-time requirements.
We use the performance monitor that comes with Windows to record the memory and single-core CPU load occupancy of the data collector during operation.
This tool can write performance data into the command window or log file to help us in subsequent processing and analysis.
In addition, we use Kafka to transmit the log data of the client to the server for detection.
During the operation of the entire detection system, we synchronously record the size of the transmitted data and the time used to calculate the bandwidth.

In addition, we also test the resource occupancy of the detection terminal to indicate the single-machine load of our detection system and the load changes for simultaneous access of multiple hosts.
This shows that our system can accept nearly a thousand hosts for detection at the same time.

Finally, we have evaluated the contribution of each module of data processing to the overall resource reduction. We likewise give the reasons for the selection of the detection model and the associated API threshold parameters, as well as the latency required for detection.

\subsection{Experiment Results}
\subsubsection{Accuracy-Overhead Balancing}\label{balance}
\begin{figure}[!h]
    \centering
    \includegraphics[width=0.45\textwidth]{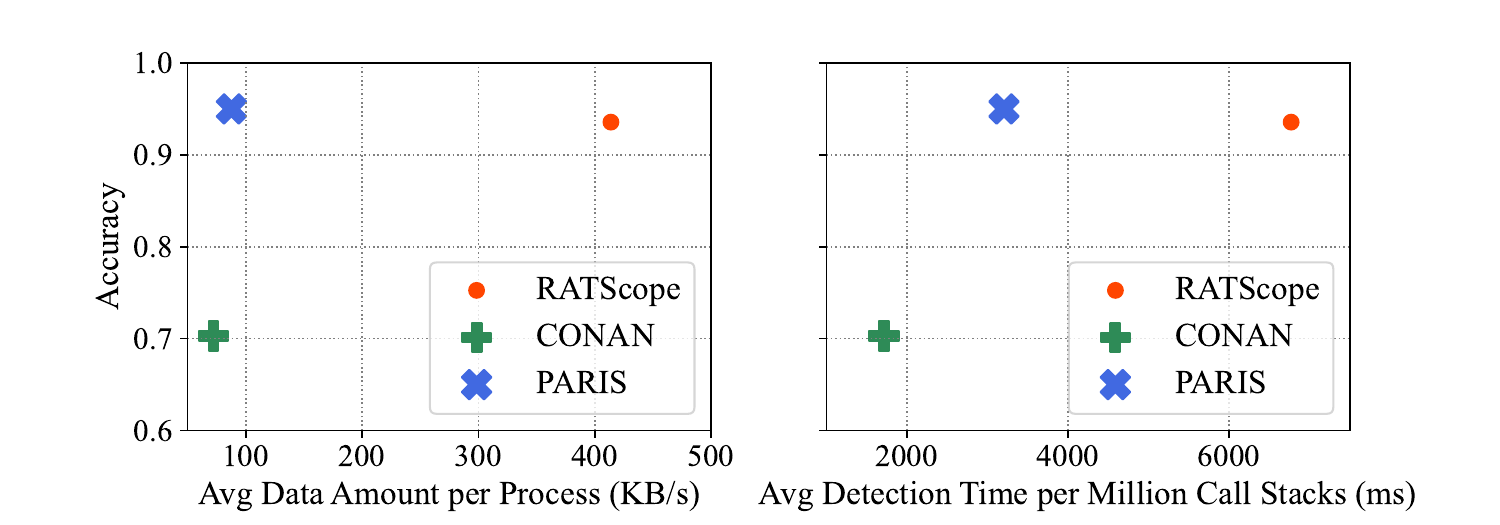}
    \caption{Comparison of overall performance between PARIS and two baselines. (includes accuracy and system load)}
    \label{fig:overall_compare}
\vspace{-0.1in}
\end{figure}

\
Fig.~\ref{fig:overall_compare} presents a comparative analysis between our method and two baselines in the malicious-behavior classification task.
From this figure, we can learn that PARIS achieves a better balance between classification accuracy, data amount, and time cost.
It is always closer to the upper left corner than the other two baselines in terms of accuracy, data amount, and data processing and detecting time.
The accuracy of PARIS and RATscope are nearly identical, with PARIS being even marginally higher at 95.00\%, compared to RATscope's 93.56\%.
This shows that we not only protect the detection capability when selectively collecting data but also improve it by eliminating some noisy and misleading API call stacks.
Furthermore, the data generation rate of PARIS is comparable to CONAN, with rates of 87.78KB/s and 72.36KB/s, respectively.
Less data typically implies lower overhead during process monitoring and data collection.
We provide a detailed overhead evaluation later in \S\ref{overhead1}.
When comparing the data processing time, we record the response time to parse and process 1 million call stacks.
The results of the PARIS fall between the two baseline methods (3204.83ms vs 6769.46ms vs 1713.54ms).
Please note that we don't consider our API-address cache module and other optimization (mentioned in \S\ref{implementation}) to get a fair result for the baselines.
In fact, if we take our cache module into account, PARIS is even much faster than CONAN (1350.95ms vs 1713.54ms), which only parses top-layer APIs.

To summarize, the experiment shows that PARIS effectively balances detection accuracy, data generation rate, and data processing time through its feature selection mechanism.

\subsubsection{Detection Result on Different Malicious Behaviors}\label{accuracy1}

\begin{table*}[h!]
\caption{The comparison of the average detection performance between our method with two different baselines.}
\vspace{0.1in}
\label{tab:PHF_table}
\centering
\renewcommand{\arraystretch}{0.9} 
\setlength{\tabcolsep}{2pt} 
\begin{tabular}{c|ccc|ccc|ccc}
\toprule
 &
   \multicolumn{3}{c|}{RATScope} &
  \multicolumn{3}{c|}{CONAN} &
  \multicolumn{3}{c}{PARIS} \\ 
 &
  \multicolumn{1}{c|}{TPR} &
  \multicolumn{1}{c|}{FPR} &
  Data Amt. &
  \multicolumn{1}{c|}{TPR} &
  \multicolumn{1}{c|}{FPR} &
  Data Amt. &
  \multicolumn{1}{c|}{TPR} &
  \multicolumn{1}{c|}{FPR} &
  Data Amt. \\ \midrule
Remote Shell&
  \multicolumn{1}{c|}{87.88\%} &
  \multicolumn{1}{c|}{3.03\%} &
  100.00\% &
  \multicolumn{1}{c|}{48.48\%} &
  \multicolumn{1}{c|}{4.30\%} &
  29.40\% &
  \multicolumn{1}{c|}{96.88\%} &
  \multicolumn{1}{c|}{5.29\%} &
  21.60\% \\ 
Keylogger &
  \multicolumn{1}{c|}{96.77\%} &
  \multicolumn{1}{c|}{10.77\%} &
  100.00\% &
  \multicolumn{1}{c|}{96.77\%} &
  \multicolumn{1}{c|}{32.75\%} &
  29.51\% &
  \multicolumn{1}{c|}{96.77\%} &
  \multicolumn{1}{c|}{5.57\%} &
  26.33\% \\ 
Desktop Capture &
  \multicolumn{1}{c|}{100.00\%} &
  \multicolumn{1}{c|}{0.49\%} &
  100.00\% &
  \multicolumn{1}{c|}{55.17\%} &
  \multicolumn{1}{c|}{0.54\%} &
  15.78\% &
  \multicolumn{1}{c|}{100.00\%} &
  \multicolumn{1}{c|}{1.19\%} &
  34.16\% \\ 
Get Clipboard&
  \multicolumn{1}{c|}{100.00\%} &
  \multicolumn{1}{c|}{3.31\%} &
  100.00\% &
  \multicolumn{1}{c|}{56.25\%} &
  \multicolumn{1}{c|}{4.57\%} &
  24.97\% &
  \multicolumn{1}{c|}{100.00\%} &
  \multicolumn{1}{c|}{0.05\%} &
  32.10\% \\ 
Open Website&
  \multicolumn{1}{c|}{70.37\%} &
  \multicolumn{1}{c|}{0.79\%} &
  100.00\% &
  \multicolumn{1}{c|}{48.15\%} &
  \multicolumn{1}{c|}{1.79\%} &
  10.48\% &
  \multicolumn{1}{c|}{70.37\%} &
  \multicolumn{1}{c|}{8.40\%} &
  50.42\% \\ 
Download and Execute&
  \multicolumn{1}{c|}{100.00\%} &
  \multicolumn{1}{c|}{5.85\%} &
  100.00\% &
  \multicolumn{1}{c|}{74.07\%} &
  \multicolumn{1}{c|}{7.35\%} &
  11.00\% &
  \multicolumn{1}{c|}{100.00\%} &
  \multicolumn{1}{c|}{4.61\%} &
  50.97\% \\ 
Audio Capture &
  \multicolumn{1}{c|}{100.00\%} &
  \multicolumn{1}{c|}{0.14\%} &
  100.00\% &
  \multicolumn{1}{c|}{66.67\%} &
  \multicolumn{1}{c|}{0.38\%} &
  45.64\% &
  \multicolumn{1}{c|}{100.00\%} &
  \multicolumn{1}{c|}{0.01\%} &
  4.85\% \\ \bottomrule
\end{tabular}
\end{table*}


Here we compare PARIS with the baselines in terms of the TPR, FPR, and Data Amount for each PHF in Table \ref{tab:PHF_table}.
To obtain stable and reliable results, we run the Random Forests, the detection model, 10 times and re-sample the data each time to get the average results.
The table clearly shows that PARIS achieves a better balance between detection accuracy and data amount than the baselines in nearly every PHF.

Compared with RATScope, PARIS gets a similar TPR in most PHF detection tasks, with a 9\% improvement in remote shell detection, and a lower FPR in 4 out of 7 tasks.
Meanwhile, the data needed for PARIS is significantly less than that for RATScope.
The result demonstrates that at least 50\% of the API data can be eliminated without compromising the detection results, which saves lots of computing resources during data collection and processing.
On the other hand, compared with CONAN, PARIS needs less data while achieving significantly better detection results for remote shells, keyloggers, and audio capturing.
Although PARIS requires more data to detect other PHFs than CONAN does, it provides more balanced results considering the large advantage in detection accuracy. 

It is noteworthy that the TPR of detecting open websites is only about 70\% in both RATScope and PARIS.
This is mainly due to the special implementation method of this PHF in some testing samples. 
In the training set, most of the RATs invoke the API \texttt{shell32.dll:ShellExecute}, which manipulates the victim's browser, to open a specified URL using the default web browser.
However, some RATs in the testing set (e.g., Imminent Monitor) opt to directly import the APIs from \texttt{ieframe.dll} to operate Internet Explorer.
This is inevitable due to the random split of the training and testing sets.
We believe we can improve the detection capability by learning from a more extensive training set.

Another reason for some slightly worse FPRs might be the inaccurate labels in the dataset.
For instance, some RATs could keep and reuse the first established remote shell alive even after we close it.
As a result, when we attempt to capture the same behavior again, the activity of "establishing a remote shell" is not actually performed, which causes some misleading labels in the dataset.
Another example is the keylogger activity, which is usually running as a background activity of the RATs.
While we are performing other malicious behaviors and collecting the execution traces, the events about the keylogger are also recorded.
The slightly poor results (TPR) for detecting opening websites and download and execution may also be caused by this.
Because some behaviors are mixed together, we may discard some useful information during API/call stack selection.

\begin{figure}[!h]
\vspace{-0.1in}
    \centering
    \includegraphics[width=0.5\textwidth]{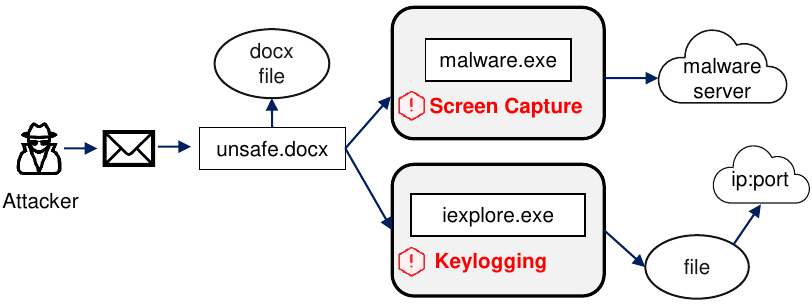}
    \caption{Attack graph generated by PARIS with the semantics of process behavior.}
    \label{fig:investigation}
\vspace{-0.2in}
\end{figure}

In addition, PARIS accurately identifies the behavioral intent of the attacker in our simulated attack experiments. As can be seen in Fig. \ref{fig:investigation}, in addition to files and processes, PARIS provides fine-grained semantic information that can help auditors better understand the attacker's strategy and intent, and respond accordingly in an emergency. For example, based on the execution time of the keylogging it is possible to determine which input information has been stolen by the attacker.

\subsubsection{Overhead Performance}\label{overhead1}
\ 
In order to evaluate the runtime load of our system, we deploy the collector and detector respectively on two real machines.
The first host (client) needs to enable the built-in ETW function and use the data collector to perform trace collection and processing. The second host (server) is responsible for receiving the trace data and performing the malicious behavior detection.
We simulate the activities of everyday users in the real world to measure system overhead.
We evaluate memory, CPU usage, and network data transmission scale, respectively.


\begin{figure*}[hbtp!]
\vspace{-0.1in}
	\centering
	\subfigure[Kafka Transmission Bandwidth.]{
    \includegraphics[width=0.3\textwidth]{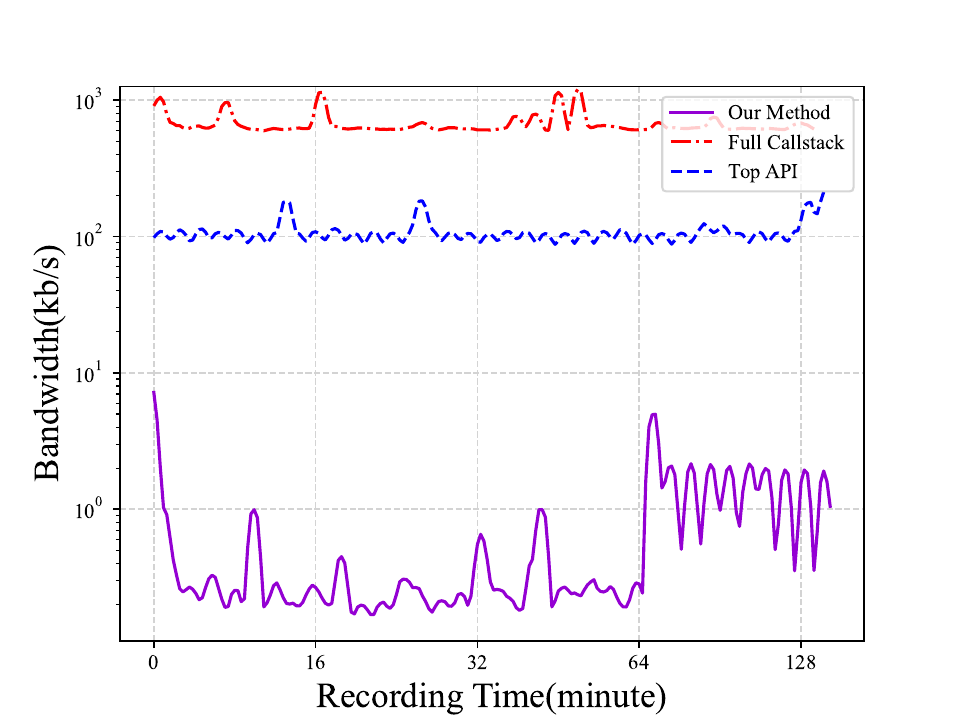}
    \label{fig:kafkadatasize}
	}
	\subfigure[Memory Usage.]{
    \includegraphics[width=0.3\textwidth]{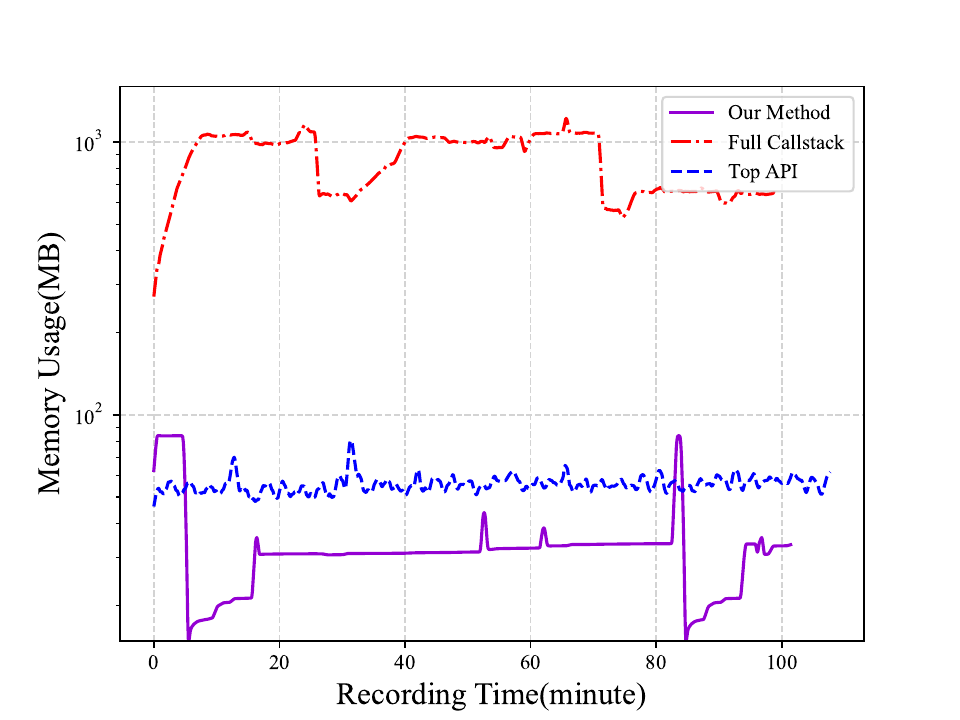}
    \label{fig:memoryuages}
	}
	\subfigure[CPU Usage.]{
    \includegraphics[width=0.3\textwidth]{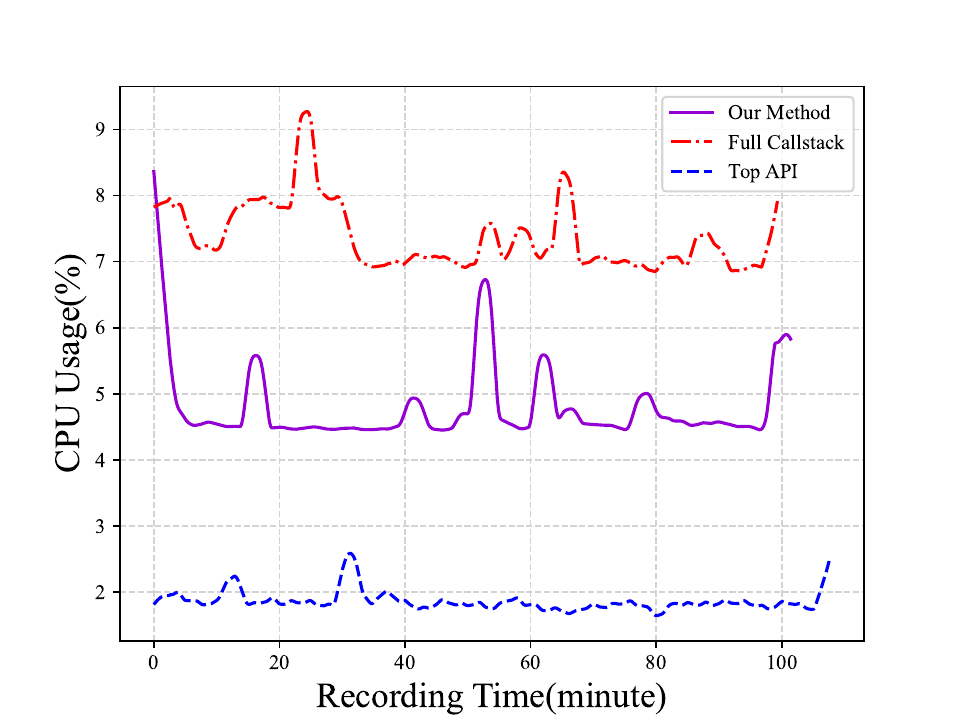}
    \label{fig:CPU}
	}	
	\caption{The system resources occupied by the data collection we deploy on the client.}
	\label{Fig: clientoverhead3image}
\end{figure*}

\textbf{Network Transmission Bandwidth.}
Fig.~\ref{fig:kafkadatasize} shows the bandwidth PARIS requires and two baselines when sending the trace to the detection server. 
The average data transmission bandwidth required by PARIS is 0.77kb/s, while the bandwidth of CONAN and RATScope are 107.24kb/s and 676.40kb/s, respectively, which are unacceptable for real-time monitoring. 
In addition, since we generate the feature vectors locally, the transmission bandwidth of the collection module is only related to the number of running processes, not the system workload.
Thus, PARIS exhibits greater bandwidth stability compared with the two baseline methods.

\textbf{Runtime Memory Usage.} Fig.~\ref{fig:memoryuages} illustrates the memory usage of PARIS and two other baselines on the client side.
The average memory usage for the collector of PARIS is 32MB, while CONAN and RATScope require an average of 58MB and 891MB of memory, respectively.
Additionally, the memory usage of PARIS depends solely on the number of running processes, so there is minimal memory fluctuation during long operations.


\textbf{Runtime CPU Usage.}
Fig.~\ref{fig:CPU} shows the CPU Usage required by PARIS and the two baselines on the client.
The results show that our method occupies an average of 4.79\% of the CPU usage, while CONAN and RATScope need to occupy 1.9\% and 7.40\% of the CPU usage, respectively.
While PARIS has a higher CPU occupancy rate than CONAN, it presents superior stability during long-term operation and a lower CPU occupancy rate than RATScope.
This is since CONAN solely parses the top-level API in each call stack, whereas PARIS employs dynamic parsing and data processing, resulting in a higher CPU usage rate.
However, the CPU occupancy of PARIS is still within the affordable range of the real-time monitoring system.



\begin{table}[h!]\footnotesize
    \caption{Overhead comparison of three detection methods under high system load environment}
    \label{tab:Stress}
\centering
\begin{tabular}{c|c|c|c}
\toprule
\textbf{Method}         & \textbf{Memory}  & \textbf{CPU} & \textbf{Bandwidth}\\ 
\midrule
RATScope & 1.495GB & 12.75\%          & 2.32MB/s\\
CONAN        & 115MB   & 4.20\%           & 563.72KB/s \\
PARIS            & 84.06MB & 12.44\%            & 14.24KB/s\\
\bottomrule
\end{tabular}
\end{table}

\begin{figure*}[hbtp!] 
\scriptsize
    \vspace{-0.1in}
	\centering
	\subfigure[Single Host]{
  \includegraphics[width=0.3\textwidth]{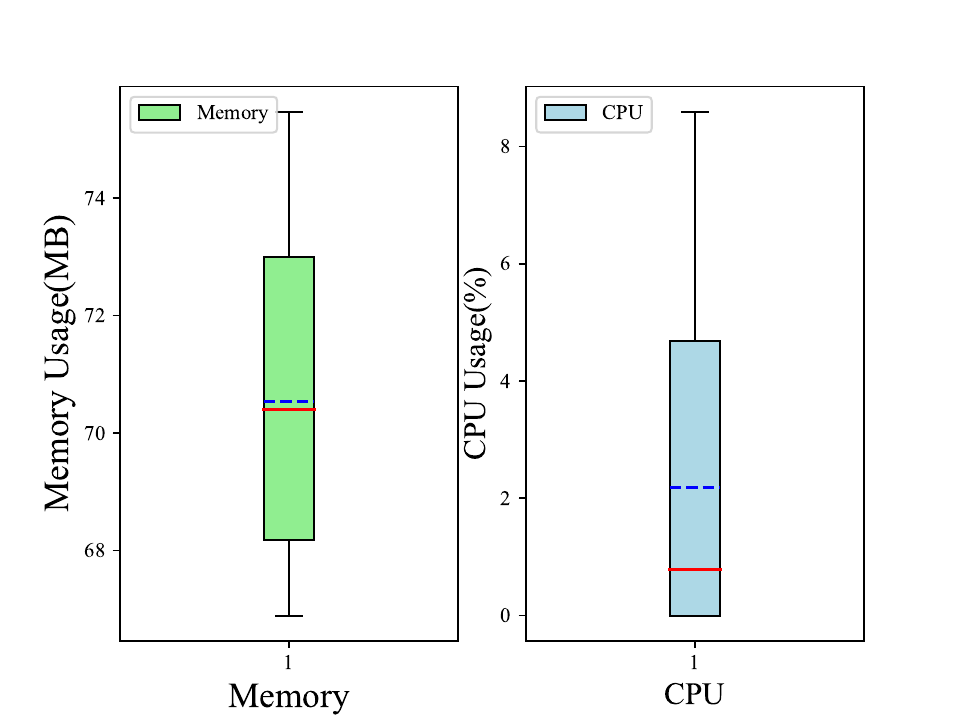}
		\label{Fig:singlehostPARIS}
	}	
	\subfigure[CPU Usage]{
  \includegraphics[width=0.3\textwidth]{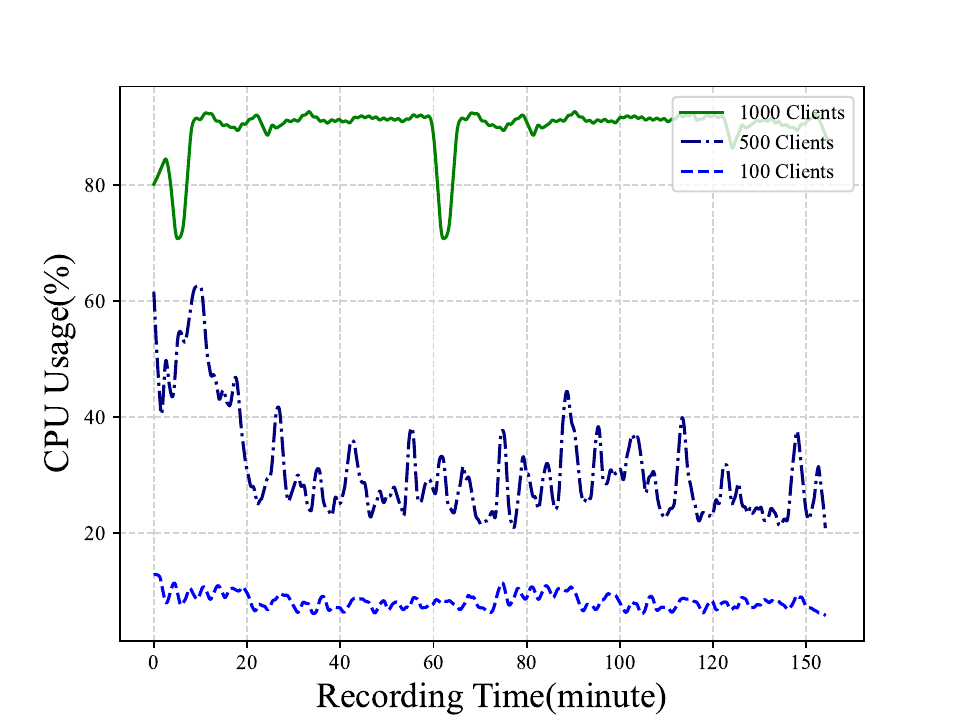}
		\label{Fig:CPUPARIS}
	}	
	\subfigure[Memory Usage]{
  \includegraphics[width=0.3\textwidth]{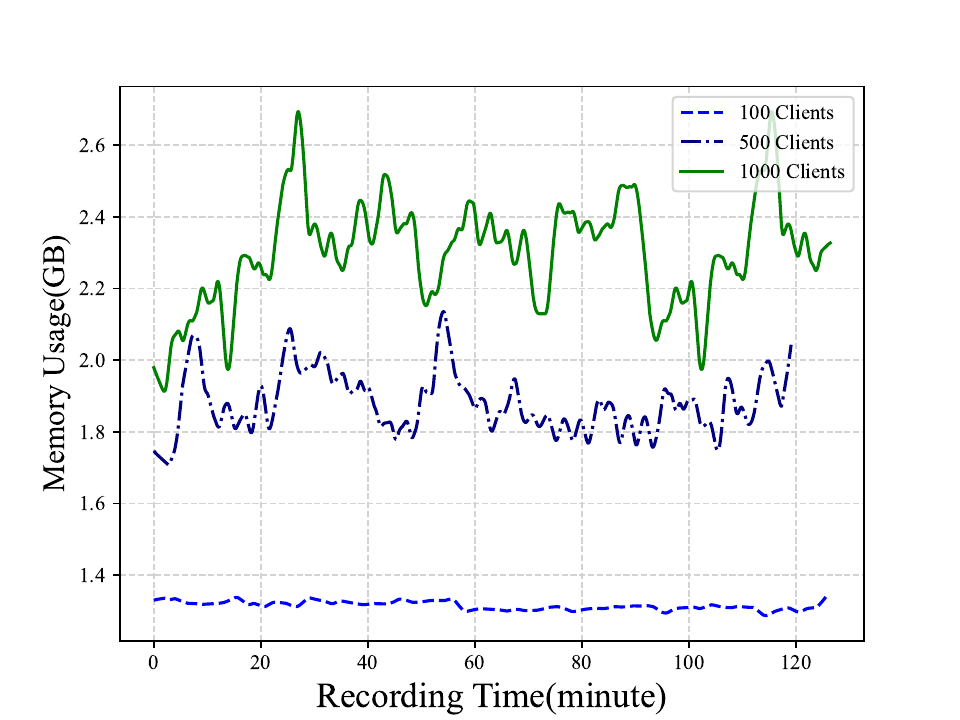}
		\label{Fig:MemoryPARIS}
	}
	\caption{The load occupancy of the detection system on our server under different number of hosts for PARIS.}
	\label{Fig: serveroverhead3image}
\end{figure*}

\begin{table}[h!]
\caption{The effect of each data processing method on data reduction}
\label{tab:Collcontribution}
\centering
\begin{tabular}{lcc}
\toprule
\textbf{Data Processing Methods} & \textbf{Remining} & \textbf{Reduce Rate} \\ 
\midrule
Raw Data                         & 100.00\%                      & -                       \\ 
Graph-based API Selection        & 62.78\%                       & -37.08\%                \\ 
Association-based API Selection & 62.58\%                       & -0.20\%                 \\ 
Call Stack Selection             & 58.39\%                     & -4.19\%                \\ 
Loop Compression                 & 31.02\%                       & -27.37\%                \\ 
Model-based API Selection   & 19.96\%                        & -11.06\%                 \\ 
Feature Extraction(API Frequency)   & 1.12\%                        & -18.84\%                 \\ \bottomrule
\end{tabular}
\end{table}

\indent \textbf{High system load.} Table \ref{tab:Stress} shows the system resource occupancy and transmission bandwidth of the collectors in the case of high system load.
The memory size of RATScope reaches nearly 1.5GB, which is not practical in a real-world system, and its CPU and bandwidth usage is also too high.
Despite CONAN's lower CPU usage compared to ours, it has a higher memory consumption and insufficient transmission bandwidth for real-time detection.
Our method takes an average of 84.06MB of system memory, 12.44\% of CPU usage, and 14.24KB/s of bandwidth.
The results show that PARIS can perform dynamic data collection and accurate attack detection with relatively low overhead in high-load situations.

\textbf{Sever Detection Overhead.} 
Fig.\ref{Fig: serveroverhead3image} shows the load occupancy of the detection system on our server under different numbers of hosts.
The server CPU used for the test is Intel(R) Xeon(R) Platinum 8272CL (8 cores, 2.60GHz operating frequency, and 32 GB memory), and CentOS Linux release 7.9.2009 (Core) server is installed. 
In this experiment, we use a simulation method to assess the resource consumption of the detector when processing system traces in a large-scale network concurrently.
As shown in Fig.~\ref{Fig:CPUPARIS} and Fig.~\ref{Fig:MemoryPARIS}, we send the trace data to the server through different numbers of topics of Kafka to simulate the consumption of data from multiple hosts at the same time.
The detector has a relatively stable memory usage, around 70 MB on average when processing data from a single host while the average CPU usage is less than 2\%.
The results of multi-client detection show that the system can monitor hundreds of hosts at the same time, and the CPU and memory usage both increase linearly.
To our knowledge, PARIS is currently the only practical system that analyzes malicious behaviors in real-time based on API call stacks and sustains continuous monitoring of a large number of hosts over extended periods~\cite{xiong2020conan}.

\subsubsection{System Efficiency}\label{other}
\
\newline
\indent \textbf{Data reduction efficiency.} Table \ref{tab:Collcontribution} shows the contributions of each step during the data collection.
As shown in the table, the average data size we finally send is only 1.12\% of the original data size.
The API selection, call stack selection, and loop compression reduced the amount of data by 80.04\% in total.
The results indicate that our data processing and feature selection steps can significantly reduce the data that needs to be sent to the detection agent, which contributes to achieving real-time and lightweight performance.

\begin{figure}[h!]
    \centering    \includegraphics[width=0.4\textwidth]{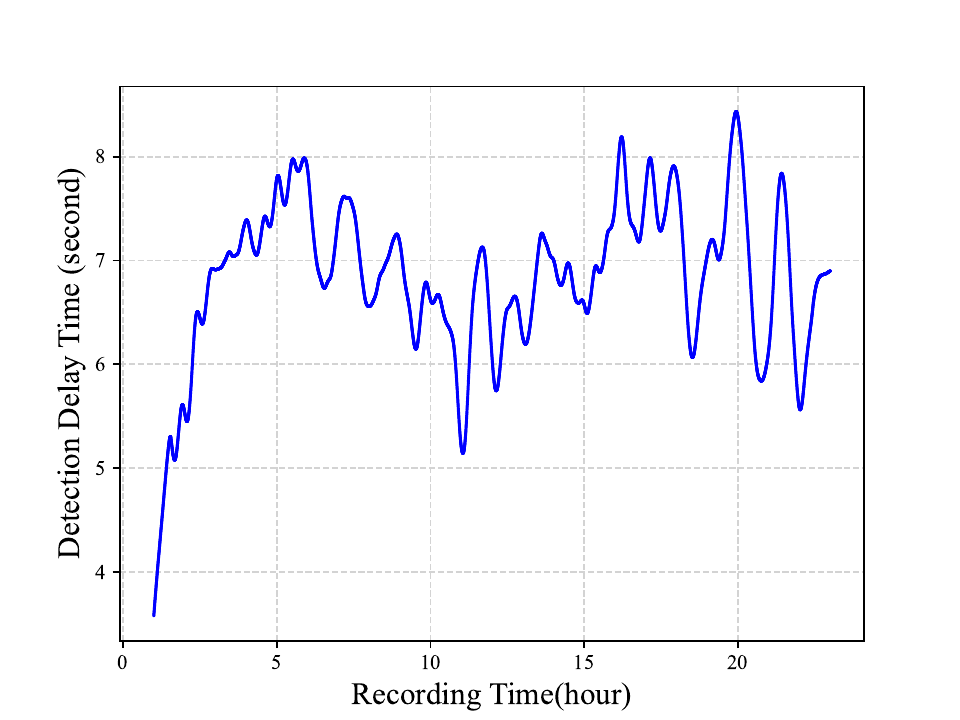}
    \caption{Detection Delay.}
    \label{fig:Detection Delay}
\end{figure}

\textbf{Detection Delay.}
We also evaluated the detection latency of PARIS. As shown in Fig.~\ref{fig:Detection Delay}, our system is able to give behavioral detection results in an average time of 6.84s, with a maximum of 10s.

\subsubsection{Ablation Study}\label{sec:ablation}
\
\newline
\indent \textbf{Select API from Models.}
We discussed selecting important APIs from the classification model in \S~\ref{sec:selectfrommodel}.
The parameters in some machine learning models can indicate the importance of each feature.
We evaluate several commonly used models that can maintain the importance of features during classification.
First, the dataset is split and utilizes 20\% of data as the validating set.
We train the classification model based on the training set and obtain the feature importances afterward.
Then, the most important parts of the features (determined by the percentiles) are kept in the validating set.
Finally, the classification accuracy is evaluated on the validating set.

\begin{figure}
    \centering    \includegraphics[width=0.35\textwidth]{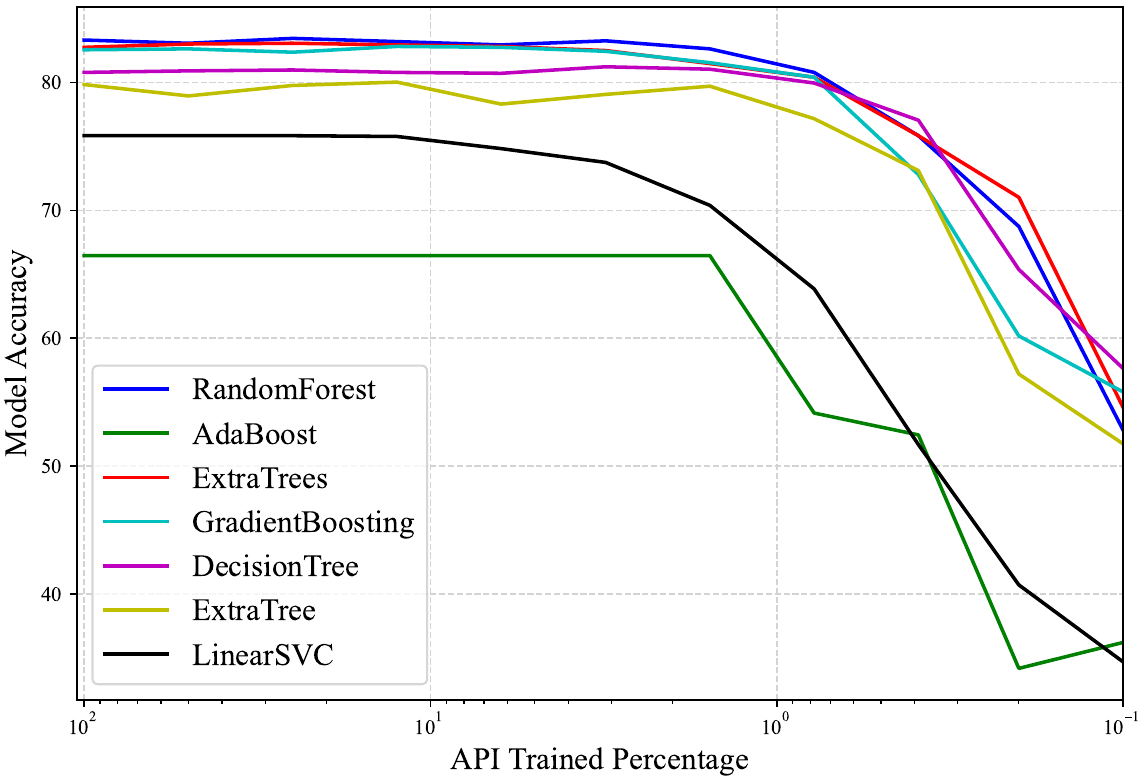}
    \caption{Model Accuracy with API Trained Percentage.}
    \label{fig:selectfrommodel}
\end{figure}

Fig.~\ref{fig:selectfrommodel} illustrates that the Random Forest achieves the best accuracy on the validating set while keeping relatively fewer APIs for detection.
It also indicates that removing over 95\% of APIs would not compromise the detection capability.
Therefore, we set the threshold as the 95th percentile of feature importance.

\begin{figure}[h!]
    \centering    \includegraphics[width=0.4\textwidth]{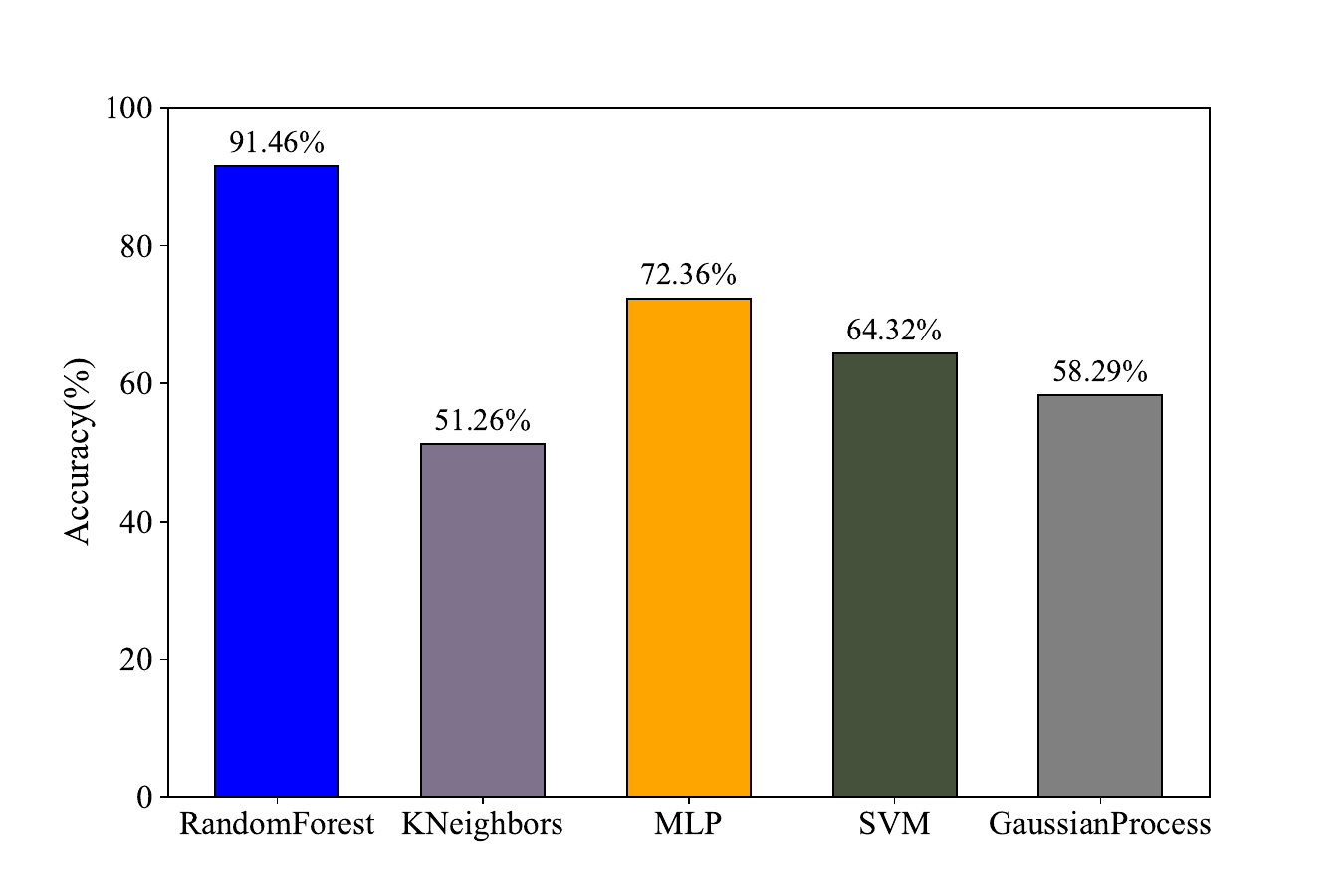}
    \vspace{-0.10in}
    \caption{Detection accuracy of different detection models.}
    \label{fig:models_accuracy}
\end{figure}

\textbf{Different Model Accuracy.}
We also evaluate the detection accuracy of different models.
The result is shown in Fig.~\ref{fig:models_accuracy}.
Since we have already evaluated some linear and tree-based models in Fig.~\ref{fig:selectfrommodel}, we only show the best one (Random Forest) and omit others in this comparison.
Among all commonly used models, Random Forest still achieves the best detection accuracy on the validating set.
This could be partially because we select APIs using Random Forest as well.


%% file: 6_discussion.tex
\section{Discussion}\label{Discussion}
\subsection{Detection Evasion}
The cornerstone of dynamic detection lies in the actual behavior a process exhibits within the system. Consequently, obfuscation and packing techniques utilized by attackers on static files, as discussed in ~\cite{dewey2012static}, will not elude our detection framework ~\cite{ucci2019survey,burnap2018malware}.
Furthermore, compared to other dynamic detection methodologies ~\cite{qiao2014cbm}, the log data we use is more fine-grained, and the features are more essential. Then, PARIS has also invested considerable effort in eliminating noise and pinpointing events associated with malicious activities, which made our model more resistant to dynamic detection bypass techniques such as random API insertion.
In summary, the PARIS system has stronger robustness to avoid the attack evasion problem.



\subsection{Model Extensibility}
The model under our investigation acquires knowledge of multiple malicious behaviors, encompassing keyloggers, remote desktops, and remote shells, among others. However, it is crucial to highlight that the applicability of our semantic restoration methodology is not confined to these behaviors alone.
Our research primarily addresses the universal challenge of inferring high-level semantics from low-level audit logs, with the detection of RAT behaviors serving as a specific application within this broader context.
Therefore, the potential extensibility of the PARIS system allows for identifying a more comprehensive range of behavioral types.

\subsection{Limitations}
As we mentioned in the \S\ref{threadmodel}, some attackers may violate our assumptions, making our model less capable of detecting those attacks.
Such attacks include the attacks against ETW and its data and the attacks that can bypass the ETW.
In recent years, many attacks have been released to disable ETW or tamper the tracing data collected by ETW~\cite{noauthor_design_2021}.
~\cite{noauthor_slingshot_2018,noauthor_lockergoga_nodate,noauthor_apt41_nodate} evaded ETW successfully by renaming extensions, running payloads or malware.

Additionally, since our machine learning model was trained based on a limited training set, an attacker can bypass our detection system by implementing malicious behavior in an unlearned way.
However, the number of API functions provided by Microsoft for the same behavior is limited; it is not easy to perform the malicious behavior differently from the standard low-level API.
We can also immediately add the unknown attack behavior to our training set, and retrain the model.

%% file: 7_related.tex
\section{Related Work}\label{RELATED}

\subsection{Machine Learning Based Malware Detection}
Numerous machine-learning methods for malware detection have emerged in recent years.
Similar to prior research, they can be classified into two categories: static detection and dynamic detection.

For static analysis, the program is examined without being executed.
Many features are used for static analysis, including binary codes, byte sequence~\cite{schultz2000data}, strings~\cite{schultz2000data,islam2010classification}, source codes, file paths~\cite{kyadige2019learning}, API calls (not retrieved from executing process)~\cite{ye2007imds}, and Opcodes~\cite{shabtai2012detecting,ye2017survey}.
Static analysis, which typically explores all execution paths of a program, can be hampered by undecidability and code obfuscation.
This often leads to the analysis being overwhelmed by a vast number of possible execution paths, making it inefficient for complex software analysis~\cite{moser2007limits,ye2017survey}.

Therefore, dynamic analysis is proposed to focus on the actual behaviors of malware.
Tracing and analyzing API calls is a significant way to infer the software behavior~\cite{chakkaravarthy2019survey,alaeiyan2019analysis,taheri2020similarity}, hence the wide use of sequence analysis techniques from the field of machine learning.
For example, Amer et al. model the software behavior via the Markov process~\cite{amer2020contextual}.
Tobiyama et al. choose the popular RNN model to analyze the sequence~\cite{tobiyama2016malware}.
Ki et al. creatively utilize multiple sequence alignment (MSA) and longest common subsequences (LCSs) in DNA analysis to study API sequences~\cite{ki2015novel}.Tran et al. use NLP-based methods to analyze the API calling sequence~\cite{tran2017nlp}.
Additionally, many other mature machine learning methods such as support vector machine and decision tree, naive Bayes classifier, and graph analysis are also suitable for this question~\cite{fan2018android,lin2018secure,fan2015malware}.

\subsection{Malicious Behavior Recognition}
At present, many researches~\cite{cono2020designing,yang2020ratscope,kwon2018mci,kolbitsch2009effective,semantics3,preda2008semantics,rabbani2020hybrid} have attempted to model and detect malware behavior semantics.
Still, the unsatisfactory detection results can be attributed to the absence of system call output parameters corresponding to the Windows system.
Besides, RATscope~\cite{yang2020ratscope} proposed a model based on API Tree Record Graph to solve the semantic detection problem, but the overhead required for graph matching is too high that it can only be analyzed offline.
CONAN~\cite{xiong2020conan} only extracts the top-level API for behavior analysis, which leads to low accuracy. PARIS proposes a new adaptive algorithm to solve this problem to achieve a balance between accuracy and overhead.

\subsection{Feature Selection in Malware Detection}
Feature selection is the process of selecting the most relevant data used for training predictive models.
It has a more important effect on malware detection due to the increasing dimensionality of datasets and the resource constraints of data collection~\cite{feizollah2015review}.
In~\cite{feizollah2015review}, Feizollah, et al. summarize four types of features in mobile malware detection but only roughly discuss the methods for feature selection.
Lin et al.~\cite{lin2015feature} extract the n-gram feature from sandbox reports and use the term frequency-inverse document frequency (TF-IDF), principal component analysis (PCA), and kernel principal component analysis (KPCA) methods to select features.
However, they only consider 187 APIs without discussing the overhead issue in online detection. 

%% file: 8_conclusion.tex
\section{Conclusion}\label{CONCLUSION}
In this paper, we propose PARIS, the first practical, adaptive trace-fetching, real-time malicious behavior detection system.
It can efficiently collect fine-grained API call stacks and accurately detect behaviors based on that without human expertise.
We are the first ones to monitor and analyze all Windows APIs (APIs defined in all DLL files under \texttt{C:\textbackslash \textbackslash Windows)} in a real-time and lightweight detection system.
We improve the runtime performance and the quality of trace data from native ETW.
On average, only 1.12\% of the original logs' size is retained, significantly reducing data transmission bandwidth, CPU usage, and memory usage.
The evaluation results show that our detection accuracy surpasses existing systems, and can operate on the client side in real-time with significantly lower system overhead.

%% file: appendix.tex
\clearpage 
\section{Appendix}
\subsection{API importance}
Table \ref{tab:APIimportance} shows the names of some APIs in our system and their corresponding importance derived from our model.

\begin{table}[]
 \caption{API importance scores extracted by our model (only some APIs are shown)}
\small
\begin{tabular}{ll}
\hline
\multicolumn{1}{c}{API}                     & \multicolumn{1}{c}{importance} \\ \hline
ntdll.dll:LdrInitializeThunk                & 0.657862256                    \\
wow64.dll:Wow64KiUserCallbackDispatcher     & 0.654900025                    \\
wow64cpu.dll:BTCpuSimulate                  & 0.654900025                    \\
wow64cpu.dll:TurboDispatchJumpAddressEnd    & 0.654900025                    \\
wow64.dll:Wow64LdrpInitialize               & 0.654406319                    \\
ntdll.dll:RtlGetAppContainerNamedObjectPath & 0.641569983                    \\
kernel32.dll:BaseThreadInitThunk            & 0.635892372                    \\
wow64.dll:Wow64SystemServiceEx              & 0.58874352                     \\
shell32.dll:OpenAs\_RunDLLA                 & 0.37274747                     \\
windows.storage.dll:SHCreateShellItemArray  & 0.349790175                    \\
shell32.dll:ILCloneFirst                    & 0.348555912                    \\
shell32.dll:ShellExecuteExW                 & 0.337447544                    \\
shell32.dll:ShellExecuteExA                 & 0.297457418                    \\
shell32.dll:ShellExecuteA                   & 0.297457418                    \\
windows.storage.dll:DllMain                 & 0.285608492                    \\
windows.storage.dll:SHCLSIDFromString       & 0.282893113                    \\
user32.dll:GetClassLongW                    & 0.277709208                    \\
user32.dll:AddClipboardFormatListener       & 0.276721797                    \\
SHCore.dll:SHCreateStreamOnFileW            & 0.275487534                    \\
ntdll.dll:ZwOpenKeyEx                       & 0.272031597                    \\
shell32.dll:StrStrW                         & 0.262157492                    \\
kernel32.dll:BaseThreadInitThunk            & 0.261910639                    \\
ntdll.dll:RtlUserThreadStart                & 0.25919526                     \\
ntdll.dll:ZwClose                           & 0.249074303                    \\
ntdll.dll:ZwQueryKey                        & 0.247593187                    \\
ntdll.dll:ZwQueryValueKey                   & 0.246358924                    \\
ntdll.dll:ZwAllocateVirtualMemory           & 0.242162429                    \\
wow64.dll:Wow64ShallowThunkSIZE\_T32TO64    & 0.240434461                    \\
wow64.dll:Wow64LogPrint                     & 0.238706492                    \\
ntdll.dll:TpCallbackIndependent             & 0.236731671                    \\
user32.dll:DispatchMessageW                 & 0.231300913                    \\
wow64.dll:Wow64AllocThreadHeap              & 0.21969884                     \\
SHCore.dll:SHCreateStreamOnFileW            & 0.210565293                    \\
shell32.dll:SHCreateItemFromParsingName     & 0.209084177                    \\
user32.dll:GetSystemMetricsForDpi           & 0.208837324                    \\
ntdll.dll:ZwQueryInformationToken           & 0.208590471                    \\
ntdll.dll:ZwProtectVirtualMemory            & 0.20562824                     \\
KernelBase.dll:CreateProcessW               & 0.202419156                    \\
shell32.dll:SHCloneSpecialIDList            & 0.202172303                    \\
ntdll.dll:KiUserCallbackDispatcher          & 0.199703777                    \\
wininet.dll:InternetReadFile                & 0.197482103                    \\
ntdll.dll:ZwSetInformationKey               & 0.194766724                    \\
kernel32.dll:RaiseInvalid16BitExeError      & 0.187607998                    \\
ntdll.dll:RtlAllocateHeap                   & 0.187114293                    \\ \hline
\end{tabular}
\label{tab:APIimportance}
\end{table}

\subsection{API Association Rules}
Table \ref{tab:APIassociation} presents a selection of API association rules extracted by our model. The table only displays a subset of the rules for clarity and brevity, indicating the complexity and depth of the data our model has analyzed.
\begin{table*}[]
\caption{API association rules extracted by our model (only some rules are shown)}
\footnotesize
\centering
\begin{tabular}{|c|c|c|c|c|}
\hline
$API_1$                                        & $API_2$                                        & support($10^{-4}$) & confidence & lift    \\ \hline
CoreMessaging.dll:CoreUICreate                & CoreMessaging.dll:CoreUICreateEx              & 23.64                           & 1.00       & 423.03  \\ \hline
CoreMessaging.dll:CoreUICreateEx              & CoreMessaging.dll:CoreUICreate                & 23.64                           & 1.00       & 423.03  \\ \hline
GdiPlus.dll:GdipDrawImageRect                 & GdiPlus.dll:GdipDrawImageRectI                & 8.54                            & 1.00       & 1170.70 \\ \hline
GdiPlus.dll:GdipDrawImageRectI                & GdiPlus.dll:GdipDrawImageRect                 & 8.54                            & 1.00       & 1170.70 \\ \hline
KernelBase.dll:CreateMutexExW                 & ntdll.dll:ZwCreateMutant                      & 15.69                           & 1.00       & 637.22  \\ \hline
KernelBase.dll:CreateProcessA                 & KernelBase.dll:CreateProcessInternalA         & 8.34                            & 1.00       & 1198.57 \\ \hline
KernelBase.dll:CreateProcessInternalA         & KernelBase.dll:CreateProcessA                 & 8.34                            & 1.00       & 1198.57 \\ \hline
TextInputFramework.dll:InputFocusChanged      & TextInputFramework.dll:TextInputClientCreate  & 42.71                           & 1.00       & 234.14  \\ \hline
TextInputFramework.dll:TextInputClientCreate  & TextInputFramework.dll:InputFocusChanged      & 42.71                           & 1.00       & 234.14  \\ \hline
advapi32.dll:CryptReleaseContext              & advapi32.dll:QueryUserServiceNameForContext   & 5.36                            & 1.00       & 1864.44 \\ \hline
advapi32.dll:ElfRegisterEventSourceW          & advapi32.dll:RegisterEventSourceW             & 19.47                           & 1.00       & 513.67  \\ \hline
advapi32.dll:QueryUserServiceNameForContext   & advapi32.dll:CryptReleaseContext              & 5.36                            & 1.00       & 1864.44 \\ \hline
advapi32.dll:RegisterEventSourceW             & advapi32.dll:ElfRegisterEventSourceW          & 19.47                           & 1.00       & 513.67  \\ \hline
gdi32.dll:GetDeviceCaps                       & win32u.dll:NtGdiGetDeviceCaps                 & 7.75                            & 1.00       & 1290.77 \\ \hline
gdi32.dll:GetTextExtentPointW                 & gdi32full.dll:GetTextExtentPointW             & 41.91                           & 1.00       & 238.58  \\ \hline
gdi32.dll:SelectObject                        & gdi32full.dll:SelectObjectImpl                & 8.54                            & 1.00       & 1170.70 \\ \hline
gdi32.dll:SetDIBits                           & gdi32full.dll:SetDIBits                       & 12.51                           & 1.00       & 799.05  \\ \hline
iertutil.dll:CreateUri                        & iertutil.dll:CreateUriPriv                    & 48.07                           & 1.00       & 208.02  \\ \hline
iertutil.dll:CreateUriPriv                    & iertutil.dll:CreateUri                        & 48.07                           & 1.00       & 208.02  \\ \hline
mswsock.dll:dn\_expand                        & ws2\_32.dll:WSAEnumProtocolsW                 & 21.26                           & 1.00       & 470.47  \\ \hline
ntdll.dll:RtlAddRefActivationContext          & ntdll.dll:TpIsTimerSet                        & 6.75                            & 1.00       & 1480.59 \\ \hline
ntdll.dll:RtlFindMessage                      & KernelBase.dll:FormatMessageW                 & 11.52                           & 1.00       & 867.93  \\ \hline
ntdll.dll:TpIsTimerSet                        & ntdll.dll:RtlAddRefActivationContext          & 6.75                            & 1.00       & 1480.59 \\ \hline
ntdll.dll:ZwCreateMutant                      & KernelBase.dll:CreateMutexExW                 & 15.69                           & 1.00       & 637.22  \\ \hline
ntdll.dll:ZwNotifyChangeKey                   & KernelBase.dll:RegNotifyChangeKeyValue        & 13.91                           & 1.00       & 719.14  \\ \hline
ntdll.dll:ZwOpenSemaphore                     & KernelBase.dll:OpenSemaphoreW                 & 11.52                           & 1.00       & 867.93  \\ \hline
ntdll.dll:ZwQueryFullAttributesFile           & KernelBase.dll:GetFileAttributesExW           & 5.96                            & 1.00       & 1678.00 \\ \hline
ntdll.dll:ZwReleaseSemaphore                  & KernelBase.dll:ReleaseSemaphore               & 6.56                            & 1.00       & 1525.45 \\ \hline
ntdll.dll:ZwUnmapViewOfSectionEx              & KernelBase.dll:UnmapViewOfFile                & 8.54                            & 1.00       & 1170.70 \\ \hline
oleaut32.dll:DispGetIDsOfNames                & oleaut32.dll:SafeArrayCreate                  & 11.52                           & 1.00       & 867.93  \\ \hline
oleaut32.dll:SafeArrayCreate                  & oleaut32.dll:DispGetIDsOfNames                & 11.52                           & 1.00       & 867.93  \\ \hline
propsys.dll:PSGetNameFromPropertyKey          & propsys.dll:PSGetPropertyDescriptionByName    & 9.73                            & 1.00       & 1027.35 \\ \hline
propsys.dll:PSGetPropertyDescriptionByName    & propsys.dll:PSGetNameFromPropertyKey          & 9.73                            & 1.00       & 1027.35 \\ \hline
user32.dll:CopyImage                          & user32.dll:CreateIconFromResourceEx           & 12.91                           & 1.00       & 774.46  \\ \hline
user32.dll:CreateIconFromResourceEx           & user32.dll:CopyImage                          & 12.91                           & 1.00       & 774.46  \\ \hline
user32.dll:DrawStateA                         & user32.dll:MessageBoxTimeoutW                 & 497.42                          & 1.00       & 20.10   \\ \hline
user32.dll:DrawStateA                         & user32.dll:MessageBoxW                        & 497.42                          & 1.00       & 20.10   \\ \hline
user32.dll:MessageBoxTimeoutW                 & user32.dll:DrawStateA                         & 497.42                          & 1.00       & 20.10   \\ \hline
user32.dll:MessageBoxTimeoutW                 & user32.dll:MessageBoxW                        & 497.42                          & 1.00       & 20.10   \\ \hline
user32.dll:MessageBoxW                        & user32.dll:DrawStateA                         & 497.42                          & 1.00       & 20.10   \\ \hline
user32.dll:MessageBoxW                        & user32.dll:MessageBoxTimeoutW                 & 497.42                          & 1.00       & 20.10   \\ \hline
win32u.dll:NtGdiGetDeviceCaps                 & gdi32.dll:GetDeviceCaps                       & 7.75                            & 1.00       & 1290.77 \\ \hline
wininet.dll:InternetSetOptionA                & wininet.dll:InternetSetOptionW                & 114.62                          & 1.00       & 87.24   \\ \hline
wininet.dll:InternetSetOptionW                & wininet.dll:InternetSetOptionA                & 114.62                          & 1.00       & 87.24   \\ \hline
winmm.dll:mciExecute                          & winmm.dll:mciSendStringA                      & 128.72                          & 1.00       & 77.69   \\ \hline
winmm.dll:mciExecute                          & winmm.dll:mciSendStringW                      & 128.72                          & 1.00       & 77.69   \\ \hline
winmm.dll:mciSendStringA                      & winmm.dll:mciExecute                          & 128.72                          & 1.00       & 77.69   \\ \hline
winmm.dll:mciSendStringA                      & winmm.dll:mciSendStringW                      & 128.72                          & 1.00       & 77.69   \\ \hline
winmm.dll:mciSendStringW                      & winmm.dll:mciExecute                          & 128.72                          & 1.00       & 77.69   \\ \hline
winmm.dll:mciSendStringW                      & winmm.dll:mciSendStringA                      & 128.72                          & 1.00       & 77.69   \\ \hline
winnsi.dll:NsiRpcRegisterChangeNotification   & winnsi.dll:NsiRpcRegisterChangeNotificationEx & 12.91                           & 1.00       & 774.46  \\ \hline
winnsi.dll:NsiRpcRegisterChangeNotificationEx & winnsi.dll:NsiRpcRegisterChangeNotification   & 12.91                           & 1.00       & 774.46  \\ \hline
\end{tabular}
\label{tab:APIassociation}
\end{table*}


%% file: Main_CCS.bbl

\begin{thebibliography}{108}


\ifx \showCODEN    \undefined \def \showCODEN     #1{\unskip}     \fi
\ifx \showDOI      \undefined \def \showDOI       #1{#1}\fi
\ifx \showISBNx    \undefined \def \showISBNx     #1{\unskip}     \fi
\ifx \showISBNxiii \undefined \def \showISBNxiii  #1{\unskip}     \fi
\ifx \showISSN     \undefined \def \showISSN      #1{\unskip}     \fi
\ifx \showLCCN     \undefined \def \showLCCN      #1{\unskip}     \fi
\ifx \shownote     \undefined \def \shownote      #1{#1}          \fi
\ifx \showarticletitle \undefined \def \showarticletitle #1{#1}   \fi
\ifx \showURL      \undefined \def \showURL       {\relax}        \fi
\providecommand\bibfield[2]{#2}
\providecommand\bibinfo[2]{#2}
\providecommand\natexlab[1]{#1}
\providecommand\showeprint[2][]{arXiv:#2}

\bibitem[Abed et~al\mbox{.}(2015)]%
        {abed2015applying}
\bibfield{author}{\bibinfo{person}{Amr~S Abed}, \bibinfo{person}{T~Charles Clancy}, {and} \bibinfo{person}{David~S Levy}.} \bibinfo{year}{2015}\natexlab{}.
\newblock \showarticletitle{Applying bag of system calls for anomalous behavior detection of applications in linux containers}. In \bibinfo{booktitle}{\emph{2015 IEEE globecom workshops (GC Wkshps)}}. IEEE, \bibinfo{pages}{1--5}.
\newblock


\bibitem[Agrawal et~al\mbox{.}(1996)]%
        {agrawal1996fast}
\bibfield{author}{\bibinfo{person}{Rakesh Agrawal}, \bibinfo{person}{Heikki Mannila}, \bibinfo{person}{Ramakrishnan Srikant}, \bibinfo{person}{Hannu Toivonen}, \bibinfo{person}{A~Inkeri Verkamo}, {et~al\mbox{.}}} \bibinfo{year}{1996}\natexlab{}.
\newblock \showarticletitle{Fast discovery of association rules.}
\newblock \bibinfo{journal}{\emph{Advances in knowledge discovery and data mining}} \bibinfo{volume}{12}, \bibinfo{number}{1} (\bibinfo{year}{1996}), \bibinfo{pages}{307--328}.
\newblock


\bibitem[Ahmed et~al\mbox{.}(2009)]%
        {ahmed2009using}
\bibfield{author}{\bibinfo{person}{Faraz Ahmed}, \bibinfo{person}{Haider Hameed}, \bibinfo{person}{M~Zubair Shafiq}, {and} \bibinfo{person}{Muddassar Farooq}.} \bibinfo{year}{2009}\natexlab{}.
\newblock \showarticletitle{Using spatio-temporal information in API calls with machine learning algorithms for malware detection}. In \bibinfo{booktitle}{\emph{Proceedings of the 2nd ACM Workshop on Security and Artificial Intelligence}}. \bibinfo{pages}{55--62}.
\newblock


\bibitem[Ahmed et~al\mbox{.}(2021a)]%
        {}
\bibfield{author}{\bibinfo{person}{Muhammad~Ejaz Ahmed}, \bibinfo{person}{Hyoungshick Kim}, \bibinfo{person}{Seyit Camtepe}, {and} \bibinfo{person}{Surya Nepal}.} \bibinfo{year}{2021}\natexlab{a}.
\newblock \showarticletitle{Peeler: Profiling kernel-level events to detect ransomware}. In \bibinfo{booktitle}{\emph{Computer Security--ESORICS 2021: 26th European Symposium on Research in Computer Security, Darmstadt, Germany, October 4--8, 2021, Proceedings, Part I 26}}. Springer, \bibinfo{pages}{240--260}.
\newblock


\bibitem[Ahmed et~al\mbox{.}(2021b)]%
        {ahmed2021peeler}
\bibfield{author}{\bibinfo{person}{Muhammad~Ejaz Ahmed}, \bibinfo{person}{Hyoungshick Kim}, \bibinfo{person}{Seyit Camtepe}, {and} \bibinfo{person}{Surya Nepal}.} \bibinfo{year}{2021}\natexlab{b}.
\newblock \showarticletitle{Peeler: Profiling Kernel-Level Events to Detect Ransomware}. In \bibinfo{booktitle}{\emph{European Symposium on Research in Computer Security}}. Springer, \bibinfo{pages}{240--260}.
\newblock


\bibitem[Ahmed et~al\mbox{.}(2020)]%
        {ahmed2020system}
\bibfield{author}{\bibinfo{person}{Yahye~Abukar Ahmed}, \bibinfo{person}{Bar{\i}{\c{s}} Ko{\c{c}}er}, \bibinfo{person}{Shamsul Huda}, \bibinfo{person}{Bander Ali~Saleh Al-rimy}, {and} \bibinfo{person}{Mohammad~Mehedi Hassan}.} \bibinfo{year}{2020}\natexlab{}.
\newblock \showarticletitle{A system call refinement-based enhanced Minimum Redundancy Maximum Relevance method for ransomware early detection}.
\newblock \bibinfo{journal}{\emph{Journal of Network and Computer Applications}}  \bibinfo{volume}{167} (\bibinfo{year}{2020}), \bibinfo{pages}{102753}.
\newblock


\bibitem[Alaeiyan et~al\mbox{.}(2019)]%
        {alaeiyan2019analysis}
\bibfield{author}{\bibinfo{person}{Mohammadhadi Alaeiyan}, \bibinfo{person}{Saeed Parsa}, {and} \bibinfo{person}{Mauro Conti}.} \bibinfo{year}{2019}\natexlab{}.
\newblock \showarticletitle{Analysis and classification of context-based malware behavior}.
\newblock \bibinfo{journal}{\emph{Computer Communications}}  \bibinfo{volume}{136} (\bibinfo{year}{2019}), \bibinfo{pages}{76--90}.
\newblock


\bibitem[Alazab et~al\mbox{.}(2011)]%
        {alazab2011zero}
\bibfield{author}{\bibinfo{person}{Mamoun Alazab}, \bibinfo{person}{Sitalakshmi Venkatraman}, \bibinfo{person}{Paul~A Watters}, \bibinfo{person}{Moutaz Alazab}, {et~al\mbox{.}}} \bibinfo{year}{2011}\natexlab{}.
\newblock \showarticletitle{Zero-day Malware Detection based on Supervised Learning Algorithms of API call Signatures.}
\newblock \bibinfo{journal}{\emph{AusDM}}  \bibinfo{volume}{11} (\bibinfo{year}{2011}), \bibinfo{pages}{171--182}.
\newblock


\bibitem[Alzaylaee et~al\mbox{.}(2020)]%
        {alzaylaee2020dl}
\bibfield{author}{\bibinfo{person}{Mohammed~K Alzaylaee}, \bibinfo{person}{Suleiman~Y Yerima}, {and} \bibinfo{person}{Sakir Sezer}.} \bibinfo{year}{2020}\natexlab{}.
\newblock \showarticletitle{DL-Droid: Deep learning based android malware detection using real devices}.
\newblock \bibinfo{journal}{\emph{Computers \& Security}}  \bibinfo{volume}{89} (\bibinfo{year}{2020}), \bibinfo{pages}{101663}.
\newblock


\bibitem[Amer et~al\mbox{.}(2020)]%
        {amer2020contextual}
\bibfield{author}{\bibinfo{person}{Eslam Amer}, \bibinfo{person}{Shaker El-Sappagh}, {and} \bibinfo{person}{Jong~Wan Hu}.} \bibinfo{year}{2020}\natexlab{}.
\newblock \showarticletitle{Contextual identification of windows malware through semantic interpretation of api call sequence}.
\newblock \bibinfo{journal}{\emph{Applied Sciences}} \bibinfo{volume}{10}, \bibinfo{number}{21} (\bibinfo{year}{2020}), \bibinfo{pages}{7673}.
\newblock


\bibitem[APT1(2012)]%
        {APT1}
APT1 \bibinfo{year}{2012}\natexlab{}.
\newblock \bibinfo{title}{Target’s Data Breach: The Commercialization of APT}.
\newblock
\newblock
\newblock
\shownote{https://goo.gl/cDYXCG}.


\bibitem[{APT41 Report}(2022)]%
        {noauthor_apt41_nodate}
{APT41 Report} \bibinfo{year}{2022}\natexlab{}.
\newblock
\newblock
\newblock
\shownote{\url{https://attack.mitre.org/groups/G0096/}}.


\bibitem[aptnotes(2020)]%
        {whitepaper}
aptnotes \bibinfo{year}{2020}\natexlab{}.
\newblock \bibinfo{title}{APT NOTES}.
\newblock
\newblock
\newblock
\shownote{\url{https://github.com/aptnotes/data/}}.


\bibitem[Bates et~al\mbox{.}(2015)]%
        {bates2015trustworthy}
\bibfield{author}{\bibinfo{person}{Adam Bates}, \bibinfo{person}{Dave~Jing Tian}, \bibinfo{person}{Kevin~RB Butler}, {and} \bibinfo{person}{Thomas Moyer}.} \bibinfo{year}{2015}\natexlab{}.
\newblock \showarticletitle{Trustworthy $\{$Whole-System$\}$ Provenance for the Linux Kernel}. In \bibinfo{booktitle}{\emph{24th USENIX Security Symposium (USENIX Security 15)}}. \bibinfo{pages}{319--334}.
\newblock


\bibitem[Bazrafshan et~al\mbox{.}(2013)]%
        {bazrafshan2013survey}
\bibfield{author}{\bibinfo{person}{Zahra Bazrafshan}, \bibinfo{person}{Hashem Hashemi}, \bibinfo{person}{Seyed Mehdi~Hazrati Fard}, {and} \bibinfo{person}{Ali Hamzeh}.} \bibinfo{year}{2013}\natexlab{}.
\newblock \showarticletitle{A survey on heuristic malware detection techniques}. In \bibinfo{booktitle}{\emph{The 5th Conference on Information and Knowledge Technology}}. IEEE, \bibinfo{pages}{113--120}.
\newblock


\bibitem[Bodon(2003)]%
        {bodon2003fast}
\bibfield{author}{\bibinfo{person}{Ferenc Bodon}.} \bibinfo{year}{2003}\natexlab{}.
\newblock \showarticletitle{A fast APRIORI implementation.}. In \bibinfo{booktitle}{\emph{FIMI}}, Vol.~\bibinfo{volume}{3}. \bibinfo{pages}{63}.
\newblock


\bibitem[Borgelt and Kruse(2002)]%
        {borgelt2002induction}
\bibfield{author}{\bibinfo{person}{Christian Borgelt} {and} \bibinfo{person}{Rudolf Kruse}.} \bibinfo{year}{2002}\natexlab{}.
\newblock \showarticletitle{Induction of association rules: Apriori implementation}. In \bibinfo{booktitle}{\emph{Compstat}}. Springer, \bibinfo{pages}{395--400}.
\newblock


\bibitem[Bose et~al\mbox{.}(2008)]%
        {bose2008behavioral}
\bibfield{author}{\bibinfo{person}{Abhijit Bose}, \bibinfo{person}{Xin Hu}, \bibinfo{person}{Kang~G Shin}, {and} \bibinfo{person}{Taejoon Park}.} \bibinfo{year}{2008}\natexlab{}.
\newblock \showarticletitle{Behavioral detection of malware on mobile handsets}. In \bibinfo{booktitle}{\emph{Proceedings of the 6th international conference on Mobile systems, applications, and services}}. \bibinfo{pages}{225--238}.
\newblock


\bibitem[Burnap et~al\mbox{.}(2018)]%
        {burnap2018malware}
\bibfield{author}{\bibinfo{person}{Pete Burnap}, \bibinfo{person}{Richard French}, \bibinfo{person}{Frederick Turner}, {and} \bibinfo{person}{Kevin Jones}.} \bibinfo{year}{2018}\natexlab{}.
\newblock \showarticletitle{Malware classification using self organising feature maps and machine activity data}.
\newblock \bibinfo{journal}{\emph{computers \& security}}  \bibinfo{volume}{73} (\bibinfo{year}{2018}), \bibinfo{pages}{399--410}.
\newblock


\bibitem[Chakkaravarthy et~al\mbox{.}(2019)]%
        {chakkaravarthy2019survey}
\bibfield{author}{\bibinfo{person}{S~Sibi Chakkaravarthy}, \bibinfo{person}{D Sangeetha}, {and} \bibinfo{person}{V Vaidehi}.} \bibinfo{year}{2019}\natexlab{}.
\newblock \showarticletitle{A survey on malware analysis and mitigation techniques}.
\newblock \bibinfo{journal}{\emph{Computer Science Review}}  \bibinfo{volume}{32} (\bibinfo{year}{2019}), \bibinfo{pages}{1--23}.
\newblock


\bibitem[Cho et~al\mbox{.}(2014)]%
        {cho2014malware}
\bibfield{author}{\bibinfo{person}{In~Kyeom Cho}, \bibinfo{person}{TaeGuen Kim}, \bibinfo{person}{Yu~Jin Shim}, \bibinfo{person}{Haeryong Park}, \bibinfo{person}{Bomin Choi}, {and} \bibinfo{person}{Eul~Gyu Im}.} \bibinfo{year}{2014}\natexlab{}.
\newblock \showarticletitle{Malware Similarity Analysis using API Sequence Alignments.}
\newblock \bibinfo{journal}{\emph{J. Internet Serv. Inf. Secur.}} \bibinfo{volume}{4}, \bibinfo{number}{4} (\bibinfo{year}{2014}), \bibinfo{pages}{103--114}.
\newblock


\bibitem[Christodorescu et~al\mbox{.}(2005)]%
        {semantics3}
\bibfield{author}{\bibinfo{person}{Mihai Christodorescu}, \bibinfo{person}{Somesh Jha}, \bibinfo{person}{Sanjit~A Seshia}, \bibinfo{person}{Dawn Song}, {and} \bibinfo{person}{Randal~E Bryant}.} \bibinfo{year}{2005}\natexlab{}.
\newblock \showarticletitle{Semantics-aware malware detection}. In \bibinfo{booktitle}{\emph{2005 IEEE symposium on security and privacy (S\&P'05)}}. IEEE, \bibinfo{pages}{32--46}.
\newblock


\bibitem[Cono~D'Elia et~al\mbox{.}(2020)]%
        {cono2020designing}
\bibfield{author}{\bibinfo{person}{Daniele Cono~D'Elia}, \bibinfo{person}{Simone Nicchi}, \bibinfo{person}{Matteo Mariani}, \bibinfo{person}{Matteo Marini}, {and} \bibinfo{person}{Federico Palmaro}.} \bibinfo{year}{2020}\natexlab{}.
\newblock \showarticletitle{Designing Robust API Monitoring Solutions}.
\newblock \bibinfo{journal}{\emph{arXiv e-prints}} (\bibinfo{year}{2020}), \bibinfo{pages}{arXiv--2005}.
\newblock


\bibitem[Corona et~al\mbox{.}(2013)]%
        {corona2013adversarial}
\bibfield{author}{\bibinfo{person}{Igino Corona}, \bibinfo{person}{Giorgio Giacinto}, {and} \bibinfo{person}{Fabio Roli}.} \bibinfo{year}{2013}\natexlab{}.
\newblock \showarticletitle{Adversarial attacks against intrusion detection systems: Taxonomy, solutions and open issues}.
\newblock \bibinfo{journal}{\emph{Information Sciences}}  \bibinfo{volume}{239} (\bibinfo{year}{2013}), \bibinfo{pages}{201--225}.
\newblock


\bibitem[Dahse and Holz(2014)]%
        {dahse2014simulation}
\bibfield{author}{\bibinfo{person}{Johannes Dahse} {and} \bibinfo{person}{Thorsten Holz}.} \bibinfo{year}{2014}\natexlab{}.
\newblock \showarticletitle{Simulation of Built-in PHP Features for Precise Static Code Analysis.}. In \bibinfo{booktitle}{\emph{NDSS}}, Vol.~\bibinfo{volume}{14}. Citeseer, \bibinfo{pages}{23--26}.
\newblock


\bibitem[darkcomet(2015)]%
        {darkcomet}
darkcomet \bibinfo{year}{2015}\natexlab{}.
\newblock \bibinfo{title}{How Hackers Are Using JeSuisCharlie To Spread Malware}.
\newblock
\newblock
\newblock
\shownote{\url{https://goo.gl/8Yjg1N}}.


\bibitem[Design Issues Of Modern EDRs:Bypassing ETW-Based Solutions(2021)]%
        {noauthor_design_2021}
Design Issues Of Modern EDRs:Bypassing ETW-Based Solutions \bibinfo{year}{2021}\natexlab{}.
\newblock
\newblock
\urldef\tempurl%
\url{https://binarly.io/posts/Design_issues_of_modern_EDRs_bypassing_ETW-based_solutions/index.html}
\showURL{%
\tempurl}


\bibitem[Dewey and Giffin(2012)]%
        {dewey2012static}
\bibfield{author}{\bibinfo{person}{David Dewey} {and} \bibinfo{person}{Jonathon~T Giffin}.} \bibinfo{year}{2012}\natexlab{}.
\newblock \showarticletitle{Static detection of C++ vtable escape vulnerabilities in binary code.}. In \bibinfo{booktitle}{\emph{NDSS}}.
\newblock


\bibitem[Ding et~al\mbox{.}(2022)]%
        {ding2022seqtrace}
\bibfield{author}{\bibinfo{person}{Zhenquan Ding}, \bibinfo{person}{Yonghe Guo}, \bibinfo{person}{Hui Xu}, \bibinfo{person}{Longchuan Yan}, \bibinfo{person}{Lei Cui}, \bibinfo{person}{Yuanlong Peng}, \bibinfo{person}{Feng Cheng}, {and} \bibinfo{person}{Zhiyu Hao}.} \bibinfo{year}{2022}\natexlab{}.
\newblock \showarticletitle{SeqTrace: API Call Tracing Based on Intel PT and VMI for Malware Detection}. In \bibinfo{booktitle}{\emph{International Conference on Algorithms and Architectures for Parallel Processing}}. Springer, \bibinfo{pages}{98--116}.
\newblock


\bibitem[ETW(2021)]%
        {etw1}
ETW \bibinfo{year}{2021}\natexlab{}.
\newblock \bibinfo{title}{Common fields in ETW events}.
\newblock
\newblock
\newblock
\shownote{\url{https://bit.ly/2zvJLDr}}.


\bibitem[ETW(2023)]%
        {ETW}
ETW \bibinfo{year}{2023}\natexlab{}.
\newblock \bibinfo{title}{Event Tracing for Windows}.
\newblock
\newblock
\newblock
\shownote{https://learn.microsoft.com/en-us/windows-hardware/drivers/devtest/event-tracing-for-windows--etw-}.


\bibitem[Fan et~al\mbox{.}(2015)]%
        {fan2015malware}
\bibfield{author}{\bibinfo{person}{Chun-I Fan}, \bibinfo{person}{Han-Wei Hsiao}, \bibinfo{person}{Chun-Han Chou}, {and} \bibinfo{person}{Yi-Fan Tseng}.} \bibinfo{year}{2015}\natexlab{}.
\newblock \showarticletitle{Malware detection systems based on API log data mining}. In \bibinfo{booktitle}{\emph{2015 IEEE 39th annual computer software and applications conference}}, Vol.~\bibinfo{volume}{3}. IEEE, \bibinfo{pages}{255--260}.
\newblock


\bibitem[Fan et~al\mbox{.}(2018)]%
        {fan2018android}
\bibfield{author}{\bibinfo{person}{Ming Fan}, \bibinfo{person}{Jun Liu}, \bibinfo{person}{Xiapu Luo}, \bibinfo{person}{Kai Chen}, \bibinfo{person}{Zhenzhou Tian}, \bibinfo{person}{Qinghua Zheng}, {and} \bibinfo{person}{Ting Liu}.} \bibinfo{year}{2018}\natexlab{}.
\newblock \showarticletitle{Android malware familial classification and representative sample selection via frequent subgraph analysis}.
\newblock \bibinfo{journal}{\emph{IEEE Transactions on Information Forensics and Security}} \bibinfo{volume}{13}, \bibinfo{number}{8} (\bibinfo{year}{2018}), \bibinfo{pages}{1890--1905}.
\newblock


\bibitem[Feizollah et~al\mbox{.}(2015)]%
        {feizollah2015review}
\bibfield{author}{\bibinfo{person}{Ali Feizollah}, \bibinfo{person}{Nor~Badrul Anuar}, \bibinfo{person}{Rosli Salleh}, {and} \bibinfo{person}{Ainuddin Wahid~Abdul Wahab}.} \bibinfo{year}{2015}\natexlab{}.
\newblock \showarticletitle{A review on feature selection in mobile malware detection}.
\newblock \bibinfo{journal}{\emph{Digital investigation}}  \bibinfo{volume}{13} (\bibinfo{year}{2015}), \bibinfo{pages}{22--37}.
\newblock


\bibitem[fireeye(2021)]%
        {fireeye}
fireeye \bibinfo{year}{2021}\natexlab{}.
\newblock \bibinfo{title}{2021 Fireeye Annual Report}.
\newblock
\newblock
\newblock
\shownote{https://bit.ly/2Ji320M}.


\bibitem[Gehani and Tariq(2012)]%
        {gehani2012spade}
\bibfield{author}{\bibinfo{person}{Ashish Gehani} {and} \bibinfo{person}{Dawood Tariq}.} \bibinfo{year}{2012}\natexlab{}.
\newblock \showarticletitle{SPADE: Support for provenance auditing in distributed environments}. In \bibinfo{booktitle}{\emph{ACM/IFIP/USENIX International Conference on Distributed Systems Platforms and Open Distributed Processing}}. Springer, \bibinfo{pages}{101--120}.
\newblock


\bibitem[hack1(2020)]%
        {hack1}
hack1 \bibinfo{year}{2020}\natexlab{}.
\newblock \bibinfo{title}{HackForums.net}.
\newblock
\newblock
\newblock
\shownote{\url{https://goo.gl/dHGFKU}}.


\bibitem[hack2(2020)]%
        {hack2}
hack2 \bibinfo{year}{2020}\natexlab{}.
\newblock \bibinfo{title}{Offensive Community}.
\newblock
\newblock
\newblock
\shownote{\url{https://goo.gl/jiFd3A}}.


\bibitem[hack3(2021)]%
        {hack3}
hack3 \bibinfo{year}{2021}\natexlab{}.
\newblock \bibinfo{title}{Hellbound Hackers}.
\newblock
\newblock
\newblock
\shownote{\url{https://goo.gl/3Xi1zg}}.


\bibitem[Han et~al\mbox{.}(2020)]%
        {han2020unicorn}
\bibfield{author}{\bibinfo{person}{Xueyuan Han}, \bibinfo{person}{Thomas Pasquier}, \bibinfo{person}{Adam Bates}, \bibinfo{person}{James Mickens}, {and} \bibinfo{person}{Margo Seltzer}.} \bibinfo{year}{2020}\natexlab{}.
\newblock \showarticletitle{Unicorn: Runtime provenance-based detector for advanced persistent threats}.
\newblock \bibinfo{journal}{\emph{arXiv preprint arXiv:2001.01525}} (\bibinfo{year}{2020}).
\newblock


\bibitem[Hassan et~al\mbox{.}(2020a)]%
        {9152771}
\bibfield{author}{\bibinfo{person}{Wajih~Ul Hassan}, \bibinfo{person}{Adam Bates}, {and} \bibinfo{person}{Daniel Marino}.} \bibinfo{year}{2020}\natexlab{a}.
\newblock \showarticletitle{Tactical Provenance Analysis for Endpoint Detection and Response Systems}. In \bibinfo{booktitle}{\emph{2020 IEEE Symposium on Security and Privacy (SP)}}. \bibinfo{pages}{1172--1189}.
\newblock
\urldef\tempurl%
\url{https://doi.org/10.1109/SP40000.2020.00096}
\showDOI{\tempurl}


\bibitem[Hassan et~al\mbox{.}(2019)]%
        {hassan2019nodoze}
\bibfield{author}{\bibinfo{person}{Wajih~Ul Hassan}, \bibinfo{person}{Shengjian Guo}, \bibinfo{person}{Ding Li}, \bibinfo{person}{Zhengzhang Chen}, \bibinfo{person}{Kangkook Jee}, \bibinfo{person}{Zhichun Li}, {and} \bibinfo{person}{Adam Bates}.} \bibinfo{year}{2019}\natexlab{}.
\newblock \showarticletitle{Nodoze: Combatting threat alert fatigue with automated provenance triage}. In \bibinfo{booktitle}{\emph{network and distributed systems security symposium}}.
\newblock


\bibitem[Hassan et~al\mbox{.}(2020b)]%
        {hassan2020omegalog}
\bibfield{author}{\bibinfo{person}{Wajih~Ul Hassan}, \bibinfo{person}{Mohammad~Ali Noureddine}, \bibinfo{person}{Pubali Datta}, {and} \bibinfo{person}{Adam Bates}.} \bibinfo{year}{2020}\natexlab{b}.
\newblock \showarticletitle{Omegalog: High-fidelity attack investigation via transparent multi-layer log analysis}. In \bibinfo{booktitle}{\emph{Network and distributed system security symposium}}.
\newblock


\bibitem[Hegland(2007)]%
        {hegland2007apriori}
\bibfield{author}{\bibinfo{person}{Markus Hegland}.} \bibinfo{year}{2007}\natexlab{}.
\newblock \showarticletitle{The apriori algorithm--a tutorial}.
\newblock \bibinfo{journal}{\emph{Mathematics and computation in imaging science and information processing}} (\bibinfo{year}{2007}), \bibinfo{pages}{209--262}.
\newblock


\bibitem[Hemalatha et~al\mbox{.}(2021)]%
        {hemalatha2021efficient}
\bibfield{author}{\bibinfo{person}{Jeyaprakash Hemalatha}, \bibinfo{person}{S~Abijah Roseline}, \bibinfo{person}{Subbiah Geetha}, \bibinfo{person}{Seifedine Kadry}, {and} \bibinfo{person}{Robertas Dama{\v{s}}evi{\v{c}}ius}.} \bibinfo{year}{2021}\natexlab{}.
\newblock \showarticletitle{An efficient densenet-based deep learning model for malware detection}.
\newblock \bibinfo{journal}{\emph{Entropy}} \bibinfo{volume}{23}, \bibinfo{number}{3} (\bibinfo{year}{2021}), \bibinfo{pages}{344}.
\newblock


\bibitem[Hsiao et~al\mbox{.}(2020)]%
        {hsiao2020hardware}
\bibfield{author}{\bibinfo{person}{Shun-Wen Hsiao}, \bibinfo{person}{Yeali~S Sun}, {and} \bibinfo{person}{Meng~Chang Chen}.} \bibinfo{year}{2020}\natexlab{}.
\newblock \showarticletitle{Hardware-assisted MMU redirection for in-guest monitoring and API profiling}.
\newblock \bibinfo{journal}{\emph{IEEE Transactions on Information Forensics and Security}}  \bibinfo{volume}{15} (\bibinfo{year}{2020}), \bibinfo{pages}{2402--2416}.
\newblock


\bibitem[Islam et~al\mbox{.}(2010)]%
        {islam2010classification}
\bibfield{author}{\bibinfo{person}{Rafiqul Islam}, \bibinfo{person}{Ronghua Tian}, \bibinfo{person}{Lynn Batten}, {and} \bibinfo{person}{Steve Versteeg}.} \bibinfo{year}{2010}\natexlab{}.
\newblock \showarticletitle{Classification of malware based on string and function feature selection}. In \bibinfo{booktitle}{\emph{2010 Second Cybercrime and Trustworthy Computing Workshop}}. IEEE, \bibinfo{pages}{9--17}.
\newblock


\bibitem[Ji et~al\mbox{.}(2016)]%
        {ji2016multi}
\bibfield{author}{\bibinfo{person}{Soo-Yeon Ji}, \bibinfo{person}{Bong-Keun Jeong}, \bibinfo{person}{Seonho Choi}, {and} \bibinfo{person}{Dong~Hyun Jeong}.} \bibinfo{year}{2016}\natexlab{}.
\newblock \showarticletitle{A multi-level intrusion detection method for abnormal network behaviors}.
\newblock \bibinfo{journal}{\emph{Journal of Network and Computer Applications}}  \bibinfo{volume}{62} (\bibinfo{year}{2016}), \bibinfo{pages}{9--17}.
\newblock


\bibitem[Kendall and McMillan(2007)]%
        {kendall2007practical}
\bibfield{author}{\bibinfo{person}{Kris Kendall} {and} \bibinfo{person}{Chad McMillan}.} \bibinfo{year}{2007}\natexlab{}.
\newblock \showarticletitle{Practical malware analysis}. In \bibinfo{booktitle}{\emph{Black Hat Conference, USA}}. \bibinfo{pages}{10}.
\newblock


\bibitem[Ki et~al\mbox{.}(2015)]%
        {ki2015novel}
\bibfield{author}{\bibinfo{person}{Youngjoon Ki}, \bibinfo{person}{Eunjin Kim}, {and} \bibinfo{person}{Huy~Kang Kim}.} \bibinfo{year}{2015}\natexlab{}.
\newblock \showarticletitle{A novel approach to detect malware based on API call sequence analysis}.
\newblock \bibinfo{journal}{\emph{International Journal of Distributed Sensor Networks}} \bibinfo{volume}{11}, \bibinfo{number}{6} (\bibinfo{year}{2015}), \bibinfo{pages}{659101}.
\newblock


\bibitem[Kim(2018)]%
        {kim2018ntmaldetect}
\bibfield{author}{\bibinfo{person}{Chan~Woo Kim}.} \bibinfo{year}{2018}\natexlab{}.
\newblock \showarticletitle{Ntmaldetect: A machine learning approach to malware detection using native api system calls}.
\newblock \bibinfo{journal}{\emph{arXiv preprint arXiv:1802.05412}} (\bibinfo{year}{2018}).
\newblock


\bibitem[King et~al\mbox{.}(2005)]%
        {king2005enriching}
\bibfield{author}{\bibinfo{person}{Samuel~T King}, \bibinfo{person}{Zhuoqing~Morley Mao}, \bibinfo{person}{Dominic~G Lucchetti}, {and} \bibinfo{person}{Peter~M Chen}.} \bibinfo{year}{2005}\natexlab{}.
\newblock \showarticletitle{Enriching Intrusion Alerts Through Multi-Host Causality.}. In \bibinfo{booktitle}{\emph{NDSS}}. Citeseer.
\newblock


\bibitem[Kolbitsch et~al\mbox{.}(2009)]%
        {kolbitsch2009effective}
\bibfield{author}{\bibinfo{person}{Clemens Kolbitsch}, \bibinfo{person}{Paolo~Milani Comparetti}, \bibinfo{person}{Christopher Kruegel}, \bibinfo{person}{Engin Kirda}, \bibinfo{person}{Xiao-yong Zhou}, {and} \bibinfo{person}{XiaoFeng Wang}.} \bibinfo{year}{2009}\natexlab{}.
\newblock \showarticletitle{Effective and efficient malware detection at the end host.}. In \bibinfo{booktitle}{\emph{USENIX security symposium}}, Vol.~\bibinfo{volume}{4}. \bibinfo{pages}{351--366}.
\newblock


\bibitem[Kwon et~al\mbox{.}(2018)]%
        {kwon2018mci}
\bibfield{author}{\bibinfo{person}{Yonghwi Kwon}, \bibinfo{person}{Fei Wang}, \bibinfo{person}{Weihang Wang}, \bibinfo{person}{Kyu~Hyung Lee}, \bibinfo{person}{Wen-Chuan Lee}, \bibinfo{person}{Shiqing Ma}, \bibinfo{person}{Xiangyu Zhang}, \bibinfo{person}{Dongyan Xu}, \bibinfo{person}{Somesh Jha}, \bibinfo{person}{Gabriela~F Ciocarlie}, {et~al\mbox{.}}} \bibinfo{year}{2018}\natexlab{}.
\newblock \showarticletitle{MCI: Modeling-based Causality Inference in Audit Logging for Attack Investigation.}. In \bibinfo{booktitle}{\emph{NDSS}}, Vol.~\bibinfo{volume}{2}. \bibinfo{pages}{4}.
\newblock


\bibitem[Kyadige et~al\mbox{.}(2019)]%
        {kyadige2019learning}
\bibfield{author}{\bibinfo{person}{Adarsh Kyadige}, \bibinfo{person}{Ethan~M Rudd}, {and} \bibinfo{person}{Konstantin Berlin}.} \bibinfo{year}{2019}\natexlab{}.
\newblock \showarticletitle{Learning from Context: Exploiting and Interpreting File Path Information for Better Malware Detection}.
\newblock \bibinfo{journal}{\emph{arXiv preprint arXiv:1905.06987}} (\bibinfo{year}{2019}).
\newblock


\bibitem[Li et~al\mbox{.}(2022)]%
        {li2022novel}
\bibfield{author}{\bibinfo{person}{Ce Li}, \bibinfo{person}{Qiujian Lv}, \bibinfo{person}{Ning Li}, \bibinfo{person}{Yan Wang}, \bibinfo{person}{Degang Sun}, {and} \bibinfo{person}{Yuanyuan Qiao}.} \bibinfo{year}{2022}\natexlab{}.
\newblock \showarticletitle{A novel deep framework for dynamic malware detection based on API sequence intrinsic features}.
\newblock \bibinfo{journal}{\emph{Computers \& Security}}  \bibinfo{volume}{116} (\bibinfo{year}{2022}), \bibinfo{pages}{102686}.
\newblock


\bibitem[Lin et~al\mbox{.}(2015)]%
        {lin2015feature}
\bibfield{author}{\bibinfo{person}{Chih-Ta Lin}, \bibinfo{person}{Nai-Jian Wang}, \bibinfo{person}{Han Xiao}, {and} \bibinfo{person}{Claudia Eckert}.} \bibinfo{year}{2015}\natexlab{}.
\newblock \showarticletitle{Feature selection and extraction for malware classification.}
\newblock \bibinfo{journal}{\emph{J. Inf. Sci. Eng.}} \bibinfo{volume}{31}, \bibinfo{number}{3} (\bibinfo{year}{2015}), \bibinfo{pages}{965--992}.
\newblock


\bibitem[Lin et~al\mbox{.}(2018)]%
        {lin2018secure}
\bibfield{author}{\bibinfo{person}{Zhaowen Lin}, \bibinfo{person}{Fei Xiao}, \bibinfo{person}{Yi Sun}, \bibinfo{person}{Yan Ma}, \bibinfo{person}{Cong-Cong Xing}, {and} \bibinfo{person}{Jun Huang}.} \bibinfo{year}{2018}\natexlab{}.
\newblock \showarticletitle{A secure encryption-based malware detection system}.
\newblock \bibinfo{journal}{\emph{KSII Transactions on Internet and Information Systems (TIIS)}} \bibinfo{volume}{12}, \bibinfo{number}{4} (\bibinfo{year}{2018}), \bibinfo{pages}{1799--1818}.
\newblock


\bibitem[Liu et~al\mbox{.}(2018)]%
        {liu2018towards}
\bibfield{author}{\bibinfo{person}{Yushan Liu}, \bibinfo{person}{Mu Zhang}, \bibinfo{person}{Ding Li}, \bibinfo{person}{Kangkook Jee}, \bibinfo{person}{Zhichun Li}, \bibinfo{person}{Zhenyu Wu}, \bibinfo{person}{Junghwan Rhee}, {and} \bibinfo{person}{Prateek Mittal}.} \bibinfo{year}{2018}\natexlab{}.
\newblock \showarticletitle{Towards a Timely Causality Analysis for Enterprise Security.}. In \bibinfo{booktitle}{\emph{NDSS}}.
\newblock


\bibitem[{LockerGoga}(2022)]%
        {noauthor_lockergoga_nodate}
{LockerGoga} \bibinfo{year}{2022}\natexlab{}.
\newblock
\newblock
\newblock
\shownote{\url{https://attack.mitre.org/software/S0372/}}.


\bibitem[Lopez et~al\mbox{.}(2017)]%
        {lopez2017survey}
\bibfield{author}{\bibinfo{person}{Juan Lopez}, \bibinfo{person}{Leonardo Babun}, \bibinfo{person}{Hidayet Aksu}, {and} \bibinfo{person}{A~Selcuk Uluagac}.} \bibinfo{year}{2017}\natexlab{}.
\newblock \showarticletitle{A survey on function and system call hooking approaches}.
\newblock \bibinfo{journal}{\emph{Journal of Hardware and Systems Security}}  \bibinfo{volume}{1} (\bibinfo{year}{2017}), \bibinfo{pages}{114--136}.
\newblock


\bibitem[Ma et~al\mbox{.}(2015)]%
        {ma2015accurate}
\bibfield{author}{\bibinfo{person}{Shiqing Ma}, \bibinfo{person}{Kyu~Hyung Lee}, \bibinfo{person}{Chung~Hwan Kim}, \bibinfo{person}{Junghwan Rhee}, \bibinfo{person}{Xiangyu Zhang}, {and} \bibinfo{person}{Dongyan Xu}.} \bibinfo{year}{2015}\natexlab{}.
\newblock \showarticletitle{Accurate, low cost and instrumentation-free security audit logging for windows}. In \bibinfo{booktitle}{\emph{Proceedings of the 31st Annual Computer Security Applications Conference}}. \bibinfo{pages}{401--410}.
\newblock


\bibitem[Malware(2016)]%
        {rat2}
Malware \bibinfo{year}{2016}\natexlab{}.
\newblock \bibinfo{title}{Adwind resurfaces, targeting Danish companies}.
\newblock
\newblock
\newblock
\shownote{\url{https://goo.gl/aJjE8J}}.


\bibitem[Malware(2019)]%
        {rat3}
Malware \bibinfo{year}{2019}\natexlab{}.
\newblock \bibinfo{title}{APT-C-27 (Goldmouse): Suspected Target Attack against the Middle East with WinRAR Exploit}.
\newblock
\newblock
\newblock
\shownote{\url{http://bit.ly/2NP3yoY}}.


\bibitem[Maniriho et~al\mbox{.}(2023)]%
        {maniriho2023api}
\bibfield{author}{\bibinfo{person}{Pascal Maniriho}, \bibinfo{person}{Abdun~Naser Mahmood}, {and} \bibinfo{person}{Mohammad Jabed~Morshed Chowdhury}.} \bibinfo{year}{2023}\natexlab{}.
\newblock \showarticletitle{API-MalDetect: Automated malware detection framework for windows based on API calls and deep learning techniques}.
\newblock \bibinfo{journal}{\emph{Journal of Network and Computer Applications}}  \bibinfo{volume}{218} (\bibinfo{year}{2023}), \bibinfo{pages}{103704}.
\newblock


\bibitem[Microsoft(2018)]%
        {cswalk}
Microsoft \bibinfo{year}{2018}\natexlab{}.
\newblock \bibinfo{title}{Stack Walking}.
\newblock
\newblock
\newblock
\shownote{https://learn.microsoft.com/en-us/previous-versions/windows/desktop/xperf/stack-walking}.


\bibitem[Milajerdi et~al\mbox{.}(2019)]%
        {milajerdi2019holmes}
\bibfield{author}{\bibinfo{person}{Sadegh~M Milajerdi}, \bibinfo{person}{Rigel Gjomemo}, \bibinfo{person}{Birhanu Eshete}, \bibinfo{person}{Ramachandran Sekar}, {and} \bibinfo{person}{VN Venkatakrishnan}.} \bibinfo{year}{2019}\natexlab{}.
\newblock \showarticletitle{Holmes: real-time apt detection through correlation of suspicious information flows}. In \bibinfo{booktitle}{\emph{2019 IEEE Symposium on Security and Privacy (SP)}}. IEEE, \bibinfo{pages}{1137--1152}.
\newblock


\bibitem[MITRE(2023)]%
        {MITRE}
MITRE \bibinfo{year}{2023}\natexlab{}.
\newblock \bibinfo{title}{MITRE ATTCK}.
\newblock
\newblock
\newblock
\shownote{\url{https://attack.mitre.org/}}.


\bibitem[Moser et~al\mbox{.}(2007)]%
        {moser2007limits}
\bibfield{author}{\bibinfo{person}{Andreas Moser}, \bibinfo{person}{Christopher Kruegel}, {and} \bibinfo{person}{Engin Kirda}.} \bibinfo{year}{2007}\natexlab{}.
\newblock \showarticletitle{Limits of static analysis for malware detection}. In \bibinfo{booktitle}{\emph{Twenty-Third Annual Computer Security Applications Conference (ACSAC 2007)}}. IEEE, \bibinfo{pages}{421--430}.
\newblock


\bibitem[{MSDN} {Library}(2022)]%
        {noauthor_msdn_nodate}
{MSDN} {Library} \bibinfo{year}{2022}\natexlab{}.
\newblock
\newblock
\newblock
\shownote{\url{https://docs.microsoft.com/en-us/windows}}.


\bibitem[Pedregosa et~al\mbox{.}(2011)]%
        {scikit-learn}
\bibfield{author}{\bibinfo{person}{F. Pedregosa}, \bibinfo{person}{G. Varoquaux}, \bibinfo{person}{A. Gramfort}, \bibinfo{person}{V. Michel}, \bibinfo{person}{B. Thirion}, \bibinfo{person}{O. Grisel}, \bibinfo{person}{M. Blondel}, \bibinfo{person}{P. Prettenhofer}, \bibinfo{person}{R. Weiss}, \bibinfo{person}{V. Dubourg}, \bibinfo{person}{J. Vanderplas}, \bibinfo{person}{A. Passos}, \bibinfo{person}{D. Cournapeau}, \bibinfo{person}{M. Brucher}, \bibinfo{person}{M. Perrot}, {and} \bibinfo{person}{E. Duchesnay}.} \bibinfo{year}{2011}\natexlab{}.
\newblock \showarticletitle{Scikit-learn: Machine Learning in {P}ython}.
\newblock \bibinfo{journal}{\emph{Journal of Machine Learning Research}}  \bibinfo{volume}{12} (\bibinfo{year}{2011}), \bibinfo{pages}{2825--2830}.
\newblock


\bibitem[Preda et~al\mbox{.}(2008)]%
        {preda2008semantics}
\bibfield{author}{\bibinfo{person}{Mila~Dalla Preda}, \bibinfo{person}{Mihai Christodorescu}, \bibinfo{person}{Somesh Jha}, {and} \bibinfo{person}{Saumya Debray}.} \bibinfo{year}{2008}\natexlab{}.
\newblock \showarticletitle{A semantics-based approach to malware detection}.
\newblock \bibinfo{journal}{\emph{ACM Transactions on Programming Languages and Systems (TOPLAS)}} \bibinfo{volume}{30}, \bibinfo{number}{5} (\bibinfo{year}{2008}), \bibinfo{pages}{1--54}.
\newblock


\bibitem[Qiao et~al\mbox{.}(2014)]%
        {qiao2014cbm}
\bibfield{author}{\bibinfo{person}{Yong Qiao}, \bibinfo{person}{Yuexiang Yang}, \bibinfo{person}{Jie He}, \bibinfo{person}{Chuan Tang}, {and} \bibinfo{person}{Zhixue Liu}.} \bibinfo{year}{2014}\natexlab{}.
\newblock \showarticletitle{CBM: free, automatic malware analysis framework using API call sequences}.
\newblock In \bibinfo{booktitle}{\emph{Knowledge engineering and management}}. \bibinfo{publisher}{Springer}, \bibinfo{pages}{225--236}.
\newblock


\bibitem[Qu et~al\mbox{.}(2016)]%
        {qu2016appshield}
\bibfield{author}{\bibinfo{person}{Zhengyang Qu}, \bibinfo{person}{Guanyu Guo}, \bibinfo{person}{Zhengyue Shao}, \bibinfo{person}{Vaibhav Rastogi}, \bibinfo{person}{Yan Chen}, \bibinfo{person}{Hao Chen}, {and} \bibinfo{person}{Wangjun Hong}.} \bibinfo{year}{2016}\natexlab{}.
\newblock \showarticletitle{Appshield: Enabling multi-entity access control cross platforms for mobile app management}. In \bibinfo{booktitle}{\emph{International Conference on Security and Privacy in Communication Systems}}. Springer, \bibinfo{pages}{3--23}.
\newblock


\bibitem[Rabbani et~al\mbox{.}(2020)]%
        {rabbani2020hybrid}
\bibfield{author}{\bibinfo{person}{Mahdi Rabbani}, \bibinfo{person}{Yong~Li Wang}, \bibinfo{person}{Reza Khoshkangini}, \bibinfo{person}{Hamed Jelodar}, \bibinfo{person}{Ruxin Zhao}, {and} \bibinfo{person}{Peng Hu}.} \bibinfo{year}{2020}\natexlab{}.
\newblock \showarticletitle{A hybrid machine learning approach for malicious behaviour detection and recognition in cloud computing}.
\newblock \bibinfo{journal}{\emph{Journal of Network and Computer Applications}}  \bibinfo{volume}{151} (\bibinfo{year}{2020}), \bibinfo{pages}{102507}.
\newblock


\bibitem[Rana et~al\mbox{.}(2022)]%
        {rana2022automated}
\bibfield{author}{\bibinfo{person}{Shubham Rana}, \bibinfo{person}{Nitesh Kumar}, \bibinfo{person}{Anand Handa}, {and} \bibinfo{person}{Sandeep~K Shukla}.} \bibinfo{year}{2022}\natexlab{}.
\newblock \showarticletitle{Automated Windows behavioral tracing for malware analysis}.
\newblock \bibinfo{journal}{\emph{Security and Privacy}} \bibinfo{volume}{5}, \bibinfo{number}{6} (\bibinfo{year}{2022}), \bibinfo{pages}{e253}.
\newblock


\bibitem[Rastogi et~al\mbox{.}(2016)]%
        {sandbox2}
\bibfield{author}{\bibinfo{person}{Vaibhav Rastogi}, \bibinfo{person}{Rui Shao}, \bibinfo{person}{Yan Chen}, \bibinfo{person}{Xiang Pan}, \bibinfo{person}{Shihong Zou}, {and} \bibinfo{person}{Ryan Riley}.} \bibinfo{year}{2016}\natexlab{}.
\newblock \showarticletitle{Detecting Hidden Attacks through the Mobile App-Web Interfaces}. In \bibinfo{booktitle}{\emph{2016 Network and Distributed System Security Symposium (NDSS)}}. The Internet.
\newblock


\bibitem[Sami et~al\mbox{.}(2010)]%
        {sami2010malware}
\bibfield{author}{\bibinfo{person}{Ashkan Sami}, \bibinfo{person}{Babak Yadegari}, \bibinfo{person}{Hossein Rahimi}, \bibinfo{person}{Naser Peiravian}, \bibinfo{person}{Sattar Hashemi}, {and} \bibinfo{person}{Ali Hamze}.} \bibinfo{year}{2010}\natexlab{}.
\newblock \showarticletitle{Malware detection based on mining API calls}. In \bibinfo{booktitle}{\emph{Proceedings of the 2010 ACM symposium on applied computing}}. \bibinfo{pages}{1020--1025}.
\newblock


\bibitem[Schultz et~al\mbox{.}(2000)]%
        {schultz2000data}
\bibfield{author}{\bibinfo{person}{Matthew~G Schultz}, \bibinfo{person}{Eleazar Eskin}, \bibinfo{person}{F Zadok}, {and} \bibinfo{person}{Salvatore~J Stolfo}.} \bibinfo{year}{2000}\natexlab{}.
\newblock \showarticletitle{Data mining methods for detection of new malicious executables}. In \bibinfo{booktitle}{\emph{Proceedings 2001 IEEE Symposium on Security and Privacy. S\&P 2001}}. IEEE, \bibinfo{pages}{38--49}.
\newblock


\bibitem[Shabtai et~al\mbox{.}(2012)]%
        {shabtai2012detecting}
\bibfield{author}{\bibinfo{person}{Asaf Shabtai}, \bibinfo{person}{Robert Moskovitch}, \bibinfo{person}{Clint Feher}, \bibinfo{person}{Shlomi Dolev}, {and} \bibinfo{person}{Yuval Elovici}.} \bibinfo{year}{2012}\natexlab{}.
\newblock \showarticletitle{Detecting unknown malicious code by applying classification techniques on opcode patterns}.
\newblock \bibinfo{journal}{\emph{Security Informatics}} \bibinfo{volume}{1}, \bibinfo{number}{1} (\bibinfo{year}{2012}), \bibinfo{pages}{1--22}.
\newblock


\bibitem[sony(2014)]%
        {sony}
sony \bibinfo{year}{2014}\natexlab{}.
\newblock \bibinfo{title}{Sony Pictures hack}.
\newblock
\newblock
\newblock
\shownote{https://goo.gl/t6oJcp}.


\bibitem[Sun et~al\mbox{.}(2006)]%
        {sun2006api}
\bibfield{author}{\bibinfo{person}{Hung-Min Sun}, \bibinfo{person}{Yue-Hsun Lin}, {and} \bibinfo{person}{Ming-Fung Wu}.} \bibinfo{year}{2006}\natexlab{}.
\newblock \showarticletitle{API monitoring system for defeating worms and exploits in MS-Windows system}. In \bibinfo{booktitle}{\emph{Information Security and Privacy: 11th Australasian Conference, ACISP 2006, Melbourne, Australia, July 3-5, 2006. Proceedings 11}}. Springer, \bibinfo{pages}{159--170}.
\newblock


\bibitem[Sung et~al\mbox{.}(2004)]%
        {sung2004static}
\bibfield{author}{\bibinfo{person}{Andrew~H Sung}, \bibinfo{person}{Jianyun Xu}, \bibinfo{person}{Patrick Chavez}, {and} \bibinfo{person}{Srinivas Mukkamala}.} \bibinfo{year}{2004}\natexlab{}.
\newblock \showarticletitle{Static analyzer of vicious executables (save)}. In \bibinfo{booktitle}{\emph{20th Annual Computer Security Applications Conference}}. IEEE, \bibinfo{pages}{326--334}.
\newblock


\bibitem[Taheri et~al\mbox{.}(2020)]%
        {taheri2020similarity}
\bibfield{author}{\bibinfo{person}{Rahim Taheri}, \bibinfo{person}{Meysam Ghahramani}, \bibinfo{person}{Reza Javidan}, \bibinfo{person}{Mohammad Shojafar}, \bibinfo{person}{Zahra Pooranian}, {and} \bibinfo{person}{Mauro Conti}.} \bibinfo{year}{2020}\natexlab{}.
\newblock \showarticletitle{Similarity-based Android malware detection using Hamming distance of static binary features}.
\newblock \bibinfo{journal}{\emph{Future Generation Computer Systems}}  \bibinfo{volume}{105} (\bibinfo{year}{2020}), \bibinfo{pages}{230--247}.
\newblock


\bibitem[The {Slingshot} {APT} {FAQ}(2018)]%
        {noauthor_slingshot_2018}
The {Slingshot} {APT} {FAQ} \bibinfo{year}{2018}\natexlab{}.
\newblock
\newblock
\newblock
\shownote{\url{https://securelist.com/apt-slingshot/84312/}}.


\bibitem[threatpost(2021)]%
        {threatpost}
threatpost \bibinfo{year}{2021}\natexlab{}.
\newblock \bibinfo{title}{Geriatric Microsoft Bug Exploited by APT Using Commodity RATs}.
\newblock
\newblock
\newblock
\shownote{https://threatpost.com/apt-commodity-rats-microsoft-bug/175601/}.


\bibitem[Tian et~al\mbox{.}(2010)]%
        {tian2010differentiating}
\bibfield{author}{\bibinfo{person}{Ronghua Tian}, \bibinfo{person}{Rafiqul Islam}, \bibinfo{person}{Lynn Batten}, {and} \bibinfo{person}{Steve Versteeg}.} \bibinfo{year}{2010}\natexlab{}.
\newblock \showarticletitle{Differentiating malware from cleanware using behavioural analysis}. In \bibinfo{booktitle}{\emph{2010 5th international conference on malicious and unwanted software}}. Ieee, \bibinfo{pages}{23--30}.
\newblock


\bibitem[Tobiyama et~al\mbox{.}(2016)]%
        {tobiyama2016malware}
\bibfield{author}{\bibinfo{person}{Shun Tobiyama}, \bibinfo{person}{Yukiko Yamaguchi}, \bibinfo{person}{Hajime Shimada}, \bibinfo{person}{Tomonori Ikuse}, {and} \bibinfo{person}{Takeshi Yagi}.} \bibinfo{year}{2016}\natexlab{}.
\newblock \showarticletitle{Malware detection with deep neural network using process behavior}. In \bibinfo{booktitle}{\emph{2016 IEEE 40th annual computer software and applications conference (COMPSAC)}}, Vol.~\bibinfo{volume}{2}. IEEE, \bibinfo{pages}{577--582}.
\newblock


\bibitem[Tran and Sato(2017)]%
        {tran2017nlp}
\bibfield{author}{\bibinfo{person}{Trung~Kien Tran} {and} \bibinfo{person}{Hiroshi Sato}.} \bibinfo{year}{2017}\natexlab{}.
\newblock \showarticletitle{NLP-based approaches for malware classification from API sequences}. In \bibinfo{booktitle}{\emph{2017 21st Asia Pacific Symposium on Intelligent and Evolutionary Systems (IES)}}. IEEE, \bibinfo{pages}{101--105}.
\newblock


\bibitem[TTPs(2022)]%
        {TTPs}
TTPs \bibinfo{year}{2022}\natexlab{}.
\newblock \bibinfo{title}{TACTICS, TECHNIQUES, AND PROCEDURES}.
\newblock
\newblock
\newblock
\shownote{\url{https://bit.ly/2Gf5T8u}}.


\bibitem[Ucci et~al\mbox{.}(2019)]%
        {ucci2019survey}
\bibfield{author}{\bibinfo{person}{Daniele Ucci}, \bibinfo{person}{Leonardo Aniello}, {and} \bibinfo{person}{Roberto Baldoni}.} \bibinfo{year}{2019}\natexlab{}.
\newblock \showarticletitle{Survey of machine learning techniques for malware analysis}.
\newblock \bibinfo{journal}{\emph{Computers \& Security}}  \bibinfo{volume}{81} (\bibinfo{year}{2019}), \bibinfo{pages}{123--147}.
\newblock


\bibitem[Venkatraman et~al\mbox{.}(2019)]%
        {venkatraman2019hybrid}
\bibfield{author}{\bibinfo{person}{Sitalakshmi Venkatraman}, \bibinfo{person}{Mamoun Alazab}, {and} \bibinfo{person}{R Vinayakumar}.} \bibinfo{year}{2019}\natexlab{}.
\newblock \showarticletitle{A hybrid deep learning image-based analysis for effective malware detection}.
\newblock \bibinfo{journal}{\emph{Journal of Information Security and Applications}}  \bibinfo{volume}{47} (\bibinfo{year}{2019}), \bibinfo{pages}{377--389}.
\newblock


\bibitem[Wan et~al\mbox{.}(2019)]%
        {wan2019practical}
\bibfield{author}{\bibinfo{person}{Zhiyuan Wan}, \bibinfo{person}{David Lo}, \bibinfo{person}{Xin Xia}, {and} \bibinfo{person}{Liang Cai}.} \bibinfo{year}{2019}\natexlab{}.
\newblock \showarticletitle{Practical and effective sandboxing for Linux containers}.
\newblock \bibinfo{journal}{\emph{Empirical Software Engineering}}  \bibinfo{volume}{24} (\bibinfo{year}{2019}), \bibinfo{pages}{4034--4070}.
\newblock


\bibitem[Wang et~al\mbox{.}(2020)]%
        {wang2020you}
\bibfield{author}{\bibinfo{person}{Qi Wang}, \bibinfo{person}{Wajih~Ul Hassan}, \bibinfo{person}{Ding Li}, \bibinfo{person}{Kangkook Jee}, \bibinfo{person}{Xiao Yu}, \bibinfo{person}{Kexuan Zou}, \bibinfo{person}{Junghwan Rhee}, \bibinfo{person}{Zhengzhang Chen}, \bibinfo{person}{Wei Cheng}, \bibinfo{person}{Carl~A Gunter}, {et~al\mbox{.}}} \bibinfo{year}{2020}\natexlab{}.
\newblock \showarticletitle{You Are What You Do: Hunting Stealthy Malware via Data Provenance Analysis.}. In \bibinfo{booktitle}{\emph{NDSS}}.
\newblock


\bibitem[Wang et~al\mbox{.}(2019)]%
        {wang2019mobile}
\bibfield{author}{\bibinfo{person}{Shanshan Wang}, \bibinfo{person}{Zhenxiang Chen}, \bibinfo{person}{Qiben Yan}, \bibinfo{person}{Bo Yang}, \bibinfo{person}{Lizhi Peng}, {and} \bibinfo{person}{Zhongtian Jia}.} \bibinfo{year}{2019}\natexlab{}.
\newblock \showarticletitle{A mobile malware detection method using behavior features in network traffic}.
\newblock \bibinfo{journal}{\emph{Journal of Network and Computer Applications}}  \bibinfo{volume}{133} (\bibinfo{year}{2019}), \bibinfo{pages}{15--25}.
\newblock


\bibitem[Wei et~al\mbox{.}(2021)]%
        {wei2021deephunter}
\bibfield{author}{\bibinfo{person}{Renzheng Wei}, \bibinfo{person}{Lijun Cai}, \bibinfo{person}{Lixin Zhao}, \bibinfo{person}{Aimin Yu}, {and} \bibinfo{person}{Dan Meng}.} \bibinfo{year}{2021}\natexlab{}.
\newblock \showarticletitle{Deephunter: A graph neural network based approach for robust cyber threat hunting}. In \bibinfo{booktitle}{\emph{Security and Privacy in Communication Networks: 17th EAI International Conference, SecureComm 2021, Virtual Event, September 6--9, 2021, Proceedings, Part I 17}}. Springer, \bibinfo{pages}{3--24}.
\newblock


\bibitem[Wong and Lie(2016)]%
        {sandbox1}
\bibfield{author}{\bibinfo{person}{Michelle~Y Wong} {and} \bibinfo{person}{David Lie}.} \bibinfo{year}{2016}\natexlab{}.
\newblock \showarticletitle{Intellidroid: a targeted input generator for the dynamic analysis of android malware.}. In \bibinfo{booktitle}{\emph{NDSS}}, Vol.~\bibinfo{volume}{16}. \bibinfo{pages}{21--24}.
\newblock


\bibitem[Xing et~al\mbox{.}(2020)]%
        {xing2020research}
\bibfield{author}{\bibinfo{person}{Jianhua Xing}, \bibinfo{person}{Hong Sheng}, \bibinfo{person}{Yuning Zheng}, {and} \bibinfo{person}{Wei Li}.} \bibinfo{year}{2020}\natexlab{}.
\newblock \showarticletitle{Research on a Malicious Code Detection Method Based on Convolutional Neural Network in a Domestic Sandbox Environment}. In \bibinfo{booktitle}{\emph{International Symposium on Cyberspace Safety and Security}}. Springer, \bibinfo{pages}{290--298}.
\newblock


\bibitem[Xiong et~al\mbox{.}(2020)]%
        {xiong2020conan}
\bibfield{author}{\bibinfo{person}{Chunlin Xiong}, \bibinfo{person}{Tiantian Zhu}, \bibinfo{person}{Weihao Dong}, \bibinfo{person}{Linqi Ruan}, \bibinfo{person}{Runqing Yang}, \bibinfo{person}{Yan Chen}, \bibinfo{person}{Yueqiang Cheng}, \bibinfo{person}{Shuai Cheng}, {and} \bibinfo{person}{Xutong Chen}.} \bibinfo{year}{2020}\natexlab{}.
\newblock \showarticletitle{CONAN: A practical real-time APT detection system with high accuracy and efficiency}.
\newblock \bibinfo{journal}{\emph{IEEE Transactions on Dependable and Secure Computing}} (\bibinfo{year}{2020}).
\newblock


\bibitem[Xtremerat(2015)]%
        {Xtremerat}
Xtremerat \bibinfo{year}{2015}\natexlab{}.
\newblock \bibinfo{title}{New Xtreme RAT Attacks US, Israel, and Other Foreign Governments}.
\newblock
\newblock
\newblock
\shownote{\url{https://goo.gl/MgmKm5}}.


\bibitem[Yang et~al\mbox{.}(2020)]%
        {yang2020ratscope}
\bibfield{author}{\bibinfo{person}{Runqing Yang}, \bibinfo{person}{Xutong Chen}, \bibinfo{person}{Haitao Xu}, \bibinfo{person}{Yueqiang Cheng}, \bibinfo{person}{Chunlin Xiong}, \bibinfo{person}{Linqi Ruan}, \bibinfo{person}{Mohammad Kavousi}, \bibinfo{person}{Zhenyuan Li}, \bibinfo{person}{Liheng Xu}, {and} \bibinfo{person}{Yan Chen}.} \bibinfo{year}{2020}\natexlab{}.
\newblock \showarticletitle{Ratscope: recording and reconstructing missing rat semantic behaviors for forensic analysis on windows}.
\newblock \bibinfo{journal}{\emph{IEEE Transactions on Dependable and Secure Computing}} (\bibinfo{year}{2020}).
\newblock


\bibitem[Ye et~al\mbox{.}(2017)]%
        {ye2017survey}
\bibfield{author}{\bibinfo{person}{Yanfang Ye}, \bibinfo{person}{Tao Li}, \bibinfo{person}{Donald Adjeroh}, {and} \bibinfo{person}{S~Sitharama Iyengar}.} \bibinfo{year}{2017}\natexlab{}.
\newblock \showarticletitle{A survey on malware detection using data mining techniques}.
\newblock \bibinfo{journal}{\emph{ACM Computing Surveys (CSUR)}} \bibinfo{volume}{50}, \bibinfo{number}{3} (\bibinfo{year}{2017}), \bibinfo{pages}{1--40}.
\newblock


\bibitem[Ye et~al\mbox{.}(2007)]%
        {ye2007imds}
\bibfield{author}{\bibinfo{person}{Yanfang Ye}, \bibinfo{person}{Dingding Wang}, \bibinfo{person}{Tao Li}, {and} \bibinfo{person}{Dongyi Ye}.} \bibinfo{year}{2007}\natexlab{}.
\newblock \showarticletitle{IMDS: Intelligent malware detection system}. In \bibinfo{booktitle}{\emph{Proceedings of the 13th ACM SIGKDD international conference on Knowledge discovery and data mining}}. \bibinfo{pages}{1043--1047}.
\newblock


\bibitem[Zeng et~al\mbox{.}(2021)]%
        {zeng2021watson}
\bibfield{author}{\bibinfo{person}{Jun Zeng}, \bibinfo{person}{Zheng~Leong Chua}, \bibinfo{person}{Yinfang Chen}, \bibinfo{person}{Kaihang Ji}, \bibinfo{person}{Zhenkai Liang}, {and} \bibinfo{person}{Jian Mao}.} \bibinfo{year}{2021}\natexlab{}.
\newblock \showarticletitle{WATSON: Abstracting Behaviors from Audit Logs via Aggregation of Contextual Semantics.}. In \bibinfo{booktitle}{\emph{NDSS}}.
\newblock


\bibitem[Zengy et~al\mbox{.}(2022)]%
        {zengy2022shadewatcher}
\bibfield{author}{\bibinfo{person}{Jun Zengy}, \bibinfo{person}{Xiang Wang}, \bibinfo{person}{Jiahao Liu}, \bibinfo{person}{Yinfang Chen}, \bibinfo{person}{Zhenkai Liang}, \bibinfo{person}{Tat-Seng Chua}, {and} \bibinfo{person}{Zheng~Leong Chua}.} \bibinfo{year}{2022}\natexlab{}.
\newblock \showarticletitle{Shadewatcher: Recommendation-guided cyber threat analysis using system audit records}. In \bibinfo{booktitle}{\emph{2022 IEEE Symposium on Security and Privacy (SP)}}. IEEE, \bibinfo{pages}{489--506}.
\newblock


\bibitem[Zhang et~al\mbox{.}(2014)]%
        {zhang2014semantics}
\bibfield{author}{\bibinfo{person}{Mu Zhang}, \bibinfo{person}{Yue Duan}, \bibinfo{person}{Heng Yin}, {and} \bibinfo{person}{Zhiruo Zhao}.} \bibinfo{year}{2014}\natexlab{}.
\newblock \showarticletitle{Semantics-aware android malware classification using weighted contextual api dependency graphs}. In \bibinfo{booktitle}{\emph{Proceedings of the 2014 ACM SIGSAC conference on computer and communications security}}. \bibinfo{pages}{1105--1116}.
\newblock


\bibitem[Zhang et~al\mbox{.}(2023)]%
        {zhang2023building}
\bibfield{author}{\bibinfo{person}{Quan Zhang}, \bibinfo{person}{Chijin Zhou}, \bibinfo{person}{Yiwen Xu}, \bibinfo{person}{Zijing Yin}, \bibinfo{person}{Mingzhe Wang}, \bibinfo{person}{Zhuo Su}, \bibinfo{person}{Chengnian Sun}, \bibinfo{person}{Yu Jiang}, {and} \bibinfo{person}{Jiaguang Sun}.} \bibinfo{year}{2023}\natexlab{}.
\newblock \showarticletitle{Building Dynamic System Call Sandbox with Partial Order Analysis}.
\newblock \bibinfo{journal}{\emph{Proceedings of the ACM on Programming Languages}} \bibinfo{volume}{7}, \bibinfo{number}{OOPSLA2} (\bibinfo{year}{2023}), \bibinfo{pages}{1253--1280}.
\newblock


\bibitem[Zhu et~al\mbox{.}(2021)]%
        {zhu2021general}
\bibfield{author}{\bibinfo{person}{Tiantian Zhu}, \bibinfo{person}{Jiayu Wang}, \bibinfo{person}{Linqi Ruan}, \bibinfo{person}{Chunlin Xiong}, \bibinfo{person}{Jinkai Yu}, \bibinfo{person}{Yaosheng Li}, \bibinfo{person}{Yan Chen}, \bibinfo{person}{Mingqi Lv}, {and} \bibinfo{person}{Tieming Chen}.} \bibinfo{year}{2021}\natexlab{}.
\newblock \showarticletitle{General, efficient, and real-time data compaction strategy for apt forensic analysis}.
\newblock \bibinfo{journal}{\emph{IEEE Transactions on Information Forensics and Security}}  \bibinfo{volume}{16} (\bibinfo{year}{2021}), \bibinfo{pages}{3312--3325}.
\newblock


\end{thebibliography}
